\documentclass[%
 floatfix,
 superscriptaddress,
aip,
 rsi,
 amsmath,amssymb,
 reprint, longbibliography,
]{revtex4-1}

\usepackage{graphicx}
\usepackage{dcolumn}
\usepackage{bm}
\usepackage[caption=false]{subfig}
\usepackage[utf8]{inputenc}
\usepackage[T1]{fontenc}
\usepackage{mathptmx}
\usepackage{etoolbox}
\usepackage[english]{babel}
\usepackage{hyperref}
\usepackage{xcolor}
\usepackage{float}
\usepackage{comment}
\usepackage{braket}

\makeatletter
\def\@email#1#2{%
 \endgroup
 \patchcmd{\titleblock@produce}
  {\frontmatter@RRAPformat}
  {\frontmatter@RRAPformat{\produce@RRAP{*#1\href{mailto:#2}{#2}}}\frontmatter@RRAPformat}
  {}{}
}%
\makeatother

\makeatletter
\let\ACME@toc@section\l@section
\let\ACME@toc@numberline\numberline@@sections
\let\l@f@appendix\l@f@section
\def\ACME@appendix@numberline#1{%
  \ACME@toc@numberline{\appendixname\ #1}%
}
\newif\ifACME@toc@numbered
\def\ACME@toc@probe@numberline#1{%
  \@if@empty{#1}{%
    \global\ACME@toc@numberedfalse
  }{%
    \global\ACME@toc@numberedtrue
  }%
}
\def\ACME@appendix@l@section#1#2{%
  \global\ACME@toc@numberedfalse
  \begingroup
    \let\Hy@toclinkstart\relax
    \let\Hy@toclinkend\relax
    \let\numberline\ACME@toc@probe@numberline
    \setbox\z@\hbox{#1}%
  \endgroup
  \ifACME@toc@numbered
    \let\numberline@@sections\ACME@appendix@numberline
    \l@@sections{}{appendix}{#1}{#2}%
    \let\numberline@@sections\ACME@toc@numberline
  \else
    \ACME@toc@section{#1}{#2}%
  \fi
}
\def\appendix@toc{\let\l@section\ACME@appendix@l@section}
\makeatother

\newcommand{\Eeff}{{E}_\mathrm{eff}}

\newcommand{\uG}{\mu \text{G}}

\newcommand{\Bin}{\mathbf{B}_{\mathrm{in}}}
\newcommand{\Bout}{\mathbf{B}_{\mathrm{out}}}
\newcommand{\nout}{\hat{\mathbf{n}}_\mathrm{out}}

\newcommand{\K}{\mathbf{K}}
\newcommand{\grad}{\mathbf{\nabla}}

\newcommand{\xhat}{\hat{\mathbf{x}}}
\newcommand{\yhat}{\hat{\mathbf{y}}}
\newcommand{\zhat}{\hat{\mathbf{z}}}

\begin{document}


\title{Compact Actively-Shielded Magnetic Field Coil within Mu-Metal Shields\\
for ACME Electric Dipole Moment Measurements}

\newcommand{\CFP}{Center for Fundamental Physics, Dept.\ of Physics and Astronomy, Northwestern University, Evanston, IL 60208, USA}

\newcommand{\Harvard}{Dept.\ of Physics, Harvard University, Cambridge, MA 02138, USA}

\newcommand{\Okayama}{Research Institute for Interdisciplinary Science, Okayama University, Okayama 700-8530, Japan}

\newcommand{\UC}{James Franck Institute and Dept.\ of Physics, University of Chicago, Chicago, IL 60637, USA}

\newcommand{\Meisenhelder}{EuQlid, College Park, MD 20740, USA}

\newcommand{\Ang}{Quantum Technology Center, University of Maryland, College Park, MD 20742, USA}

\newcommand{\Lascar}{Rockwell Automation, Chicago, IL}

\newcommand{\CUA}{Harvard-MIT Center for Ultracold Atoms, Cambridge, MA 02138, USA}

\newcommand{\Wu}{Facility for Rare Isotope Beams, Michigan State University, East Lansing, MI 48824, USA}

\author{S. Liu}
\affiliation{\CFP} 
\author{M. Watts}
\affiliation{\CFP} 
\author{C. Diver}
\affiliation{\CFP}
\author{D. G. Ang}
\altaffiliation[Current address: ]{\Ang}
\affiliation{\CFP}
\affiliation{\Harvard}

\author{C. Meisenhelder}
\altaffiliation[Current address: ]{\Meisenhelder}
\affiliation{\CFP}
\affiliation{\Harvard}
\author{X. Fan}
\affiliation{\CFP}
\affiliation{\Harvard}
\affiliation{\CUA}
 
\author{B. Hao}
\affiliation{\CFP} 
\author{D. Lascar}
\altaffiliation{\Lascar}
\affiliation{\CFP}

\author{A. Hiramoto}
\affiliation{\CFP}
\affiliation{\Okayama}
\author{T. Masuda}
\affiliation{\Okayama}

\author{P. Hu}
\affiliation{\UC} 
\author{Z. Han}
\affiliation{\UC} 
\author{X. Wu}
\affiliation{\UC}
\altaffiliation[Current address: ]{\Wu}

\author{D. DeMille}
\affiliation{\UC} 

\author{J.M. Doyle}
\affiliation{\Harvard}
\affiliation{\CUA}

\author{G. Gabrielse$^{*,}$}
\email{gerald.gabrielse@northwestern.edu}
\affiliation{\CFP} 

\date{\today}

\begin{abstract}

A system of actively-shielded coils and mu-metal shields is devised, constructed and shown to provide the stable and spatially uniform magnetic field needed for the ACME III electron electric dipole moment (eEDM) measurement.  Two layers of current-carrying coils, enclosed within three layers of ferromagnetic shields, produce a field that varies by less than 1 nT (10 $\uG$) within the 1 m $\times$ 4.2 cm $\times$ 4.2 cm interior volume in which a beam of ThO molecules are probed as they precess. The demountable shields are constructed from rectangular mu-metal plates. The largest, with a mass of 19 kg and an area of 2.18 m $\times$ 0.75 m, is easily carried by two people and just fits within a large available annealing oven. The assembly design facilitates low-stress mounting and handling to suppress changes in the magnetic properties of the mu metal, and also provides modular access to apparatus within the coils for maintenance and upgrades. The nearly static external ambient field is reduced within the shielded precession volume by up to a factor of $10^5$. During the magnetic field reversals that ACME uses to suppress systematic uncertainties, the ``actively-shielded'' coil largely cancels out its external fringing field to minimize the magnetization of the mu metal.  Even though the shields are only 10 cm outside the coils, shield degaussing after every magnetic field reversal is not required.  The non-reversing residual field stays below 1 nT for up to 17 hours when the field is reversed every 30 seconds, for example. The measured performance, compared to the previous generation ACME II apparatus, suggests that the magnetic-field-related systematic uncertainties for ACME III will be smaller by an estimated factor of 40 despite a five times longer precession volume and the use of three magnetic shielding layers rather than five.
\end{abstract}

\maketitle

\newpage
\tableofcontents

\section{Overview}

The ACME I \cite{AcmeEdm2014,AcmeEdmDetails2016} and ACME II \cite{AcmeEdm2018} measurements of the electron's electric dipole moment each improved the sensitivity to an eEDM by an order of magnitude over its predecessor. Subsequently, a JILA measurement increased the eEDM sensitivity by approximately a factor of two \cite{jilaedm2023}. Many theoretical analyses (e.g.\  \cite{Reece2017,Reece2019,Baryogenesis2021,FlavorProbes2022}) used these eEDM bounds to test the predictions of proposed extensions to the Standard Model of Particle Physics (SM).  The experimental limits showed that some previously-favored models were untenable or required fine-tuning.  New constraints were set upon models and interactions that are beyond the Standard Model (BSM).    

Following the ACME II measurement, several steps prepared the way for a much more sensitive ACME III measurement.  The shot noise limit (not reached in ACME II) was attained \cite{AcmeShotNoise2019}.  A measurement of the lifetime of the $H$ state in Thorium monoxide (ThO) showed that a five times longer coherence time could be realized \cite{AcmeLifetime2022}.  An electrostatic lens \cite{AcmeLens2022,Wu2020} captured more molecules from a laser ablation source.  Upgraded collection optics and silicon photomultiplier (SiPM) detectors with a higher quantum efficiency were demonstrated \cite{AcmeSipm2023,masudasipm,Hiramoto2022sipm,AngThesis2023}. Improvements in the ThO cryogenic buffer-gas beam source allow for further improvement \cite{hanCryogenicBufferGas2026}. If the systematic uncertainties can be sufficiently reduced, it seems feasible to carry out an ACME III measurement that either detects the eEDM for the first time, or sets a limit up to 40 times smaller than ACME II \cite{AcmeLifetime2022,AngThesis2023,MeisenhelderThesis2023}. ACME III would thus have a sensitivity 10 times better than the recent JILA/NIST measurement, which reported \cite{jilaedm2023} a sensitivity 2.4 times that of ACME II.

\begin{figure}[htbp!]
    \centering
\includegraphics[width=\linewidth]{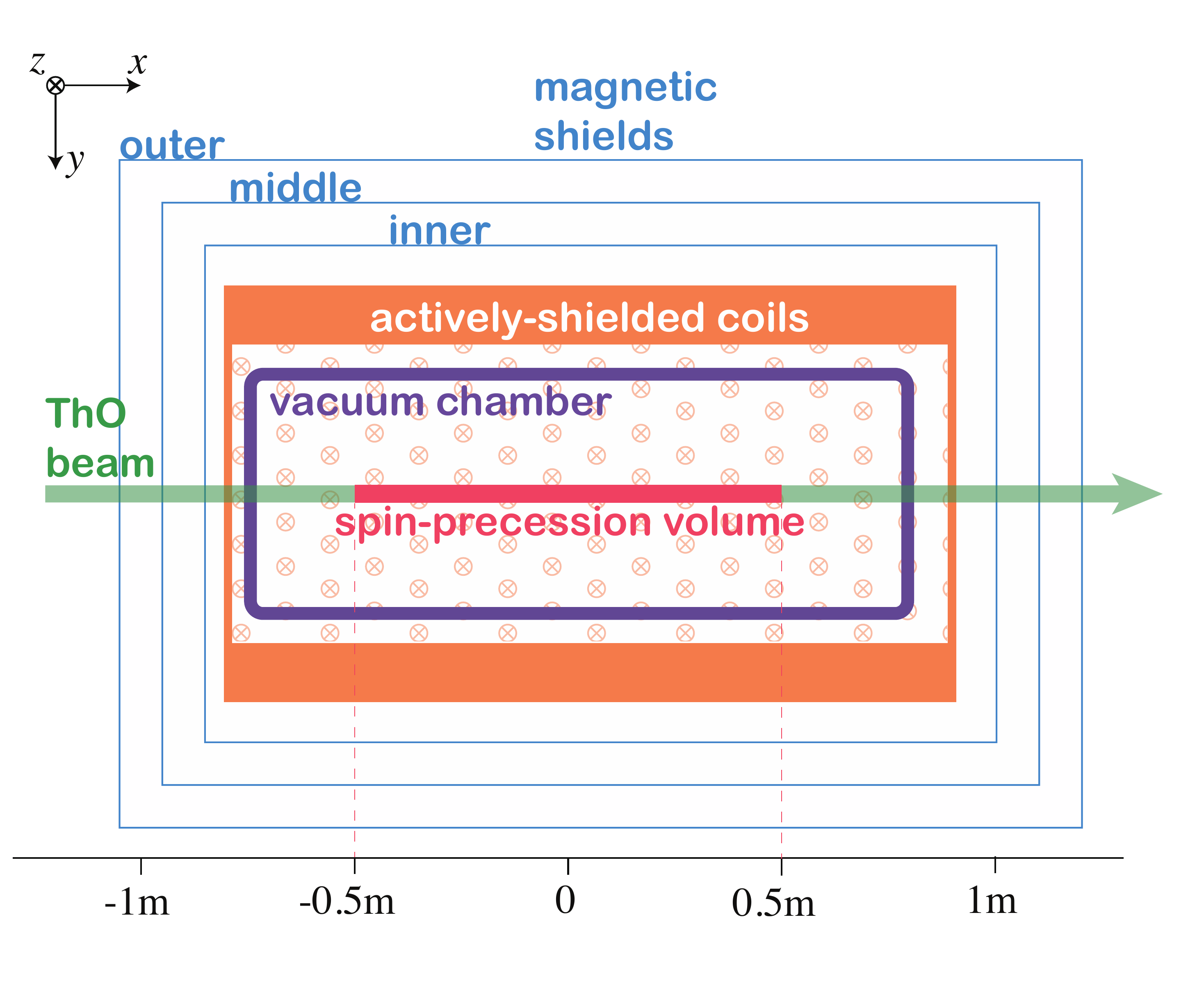} 
     \caption{Overview of the coil-plus-shields system in the context of the ACME experiment. A beam of ThO molecules (green) travels through a spin-precession volume (red) inside a vacuum chamber (purple) within the actively-shielded coil (orange). Three layers of mu-metal shielding (blue) enclose everything.}
    \label{fig:Overview}
\end{figure}

A magnetic field that varies during the spin precession of ThO molecules causes a systematic uncertainty due to the diffusion of the quantum phase measured to determine the eEDM.  The advance reported in this work is the realization of a coil-plus-mu-metal-shields system that produces the field in a long 1 m $\times$ 4.2 cm $\times$ 4.2 cm precession volume that is needed to achieve the ACME III measurement goal.  As represented in Fig.~\ref{fig:Overview}, the precession volume is contained within a large vacuum chamber. A two-layer, actively-shielded coil system encloses the vacuum chamber.  Three rectangular layers of mu-metal shields enclose the coil system.  The various layers are separated by only about 10 cm to keep the system as compact and the mass of the mu-metal as small as possible. When the coil produces the magnetic field needed for ACME III, we demonstrate that the magnetic field from the coil and the residual magnetic field within the shielded volume together vary by less than 1 nT over the long precession volume.   

In ACME I, ACME II, and similar measurements, a significant amount of time was spent degaussing the mu-metal shields every time the magnitude or direction of the magnetic field produced in the shielded region was changed. Despite having only a 10 cm separation between the field-producing coils and the nearest mu-metal shield, the requirement for frequent degaussing has largely been eliminated by our novel active shielding coil design. The ``actively-shielded'' coil is designed so that the current in its own windings largely cancels the fringing field outside the coil at the location of the nearest mu-metal shield. (This is different from a ``self-shielded'' coil system that uses flux conservation to substantially cancel temporal fluctuations of a magnetic field \cite{Gabrielse1988}.) We demonstrate that the shields are not significantly magnetized when a magnetic field of the strength planned for ACME III is reversed in direction every 30 seconds for more than a day. The coil system is also constructed so that residual magnetic fields and gradients of the shields can be canceled if they arise.

 Containing the interaction volume and a field coil far inside a large room shielded by 7 layers of mu-metal (reported in \cite{neutronEDMShield2015TUM}) was precluded by the available university space and budget.  In our compact design, successive layers of the chamber, coils, and shields are separated by only $\sim$10 cm. To minimize the weight, size, and cost, only three layers of mu-metal shielding are used.

The requirements for a coil-plus-shields system for eEDM measurements are discussed in Sec.\ \ref{sec:ACMEIII}. Of particular importance is avoiding residual magnetism of the shields, which produces a field that does not reverse direction when the current in the coils is reversed. For ACME II, such magnetism built up over time as magnetic fields inside the shields were frequently reversed. 

The actively-shielded coil that makes it possible to realize a very small non-reversing field, despite coil windings being only 10 cm from the nearest magnetic shield, is discussed in Sec.\ \ref{sec:Coil}. The coil produces the needed uniform magnetic field in the precession volume. However, the ratio of the fringing field at the nearest shield, where magnetization could occur, to the field at the center, is 20 times smaller than in ACME II, reducing the chance of shield magnetization.   

The magnetic shields that isolate the applied field from the ambient field outside the shields are described in Sec.\ \ref{sec:MagneticShields}.  The shields are made from the largest mu-metal plates that can be suspended in the largest annealing oven which we could access.  Each plate can be safely handled by two people and can be readily transported.  The shields can be fully demounted and reassembled without subjecting them to an appreciable mechanical stress that would impair their shielding performance\cite{Preece1971}.  The shields, after being degaussed using 108 degaussing coils wound around them, reduced the ambient field by about $10^5$, and produced a residual, non-reversing field no larger than $\sim5 \, \uG$.

The measured performance of the fully assembled, compact coil-plus-shields system (Fig.\ \ref{fig:Overview}) that opens the way to an ACME III measurement with 40 times better sensitivity than ACME II is described in Sec.\ \ref{sec:Performance}. The large aluminum vacuum chamber was either empty or removed for these measurements. Field plates, collection optics, and other additions within the vacuum chamber that are required to perform an actual eEDM measurement were added and the shield was reassembled. These additions, along with mechanical stress from reassembly, increased the residual field slightly, but only by an amount that could be shimmed out by adjusting the current in the coils. App.~\ref{sec:AuxCoils} briefly summarizes the design and performance of auxiliary coil windings added in order to allow a search for gradient-induced systematic errors, and also to provide a way to correct for residual field gradients if needed.


\section{\label{sec:ACMEIII} ACME Method and ACME III Requirements}

\subsection{The ACME Measurements}
\label{sec:ACME}

For an ACME measurement, a prepared superposition of two states in the $H$~$^3\Delta_1$ manifold of ThO states travels in the $\xhat$ direction through a precession volume. (Fig.\ \ref{fig:Overview} illustrates this for the ACME III apparatus reported here.)   An electric field $\mathbf{E}=\Eeff \zhat$ and a magnetic field  $\mathbf{B}=B_z\,\zhat$ are applied in the perpendicular $\zhat$ direction.   The extremely large effective field for valence electrons in the $H$-state is calculated to be $\Eeff = 78$ GV/cm \cite{Denis2016eeff, Skripnikov2016}.  Its direction can be reversed by reversing a laboratory electric field of only a few tens of V/cm \cite{PbO3Delta1Proposal2001}.   

ACME measures the change in the relative phase of the two states,
\begin{equation}
\phi(\Eeff,B_z) =  (d_e \Eeff + \mu B_z )\tau/\hbar,
\label{eq:Phase}
\end{equation}
as the superposition precesses for a time, $\tau$, through the length of the precession volume.   Here $d_e$ is the eEDM we seek.  The $H$-state magnetic moment, $\mu = -0.004 ~\mu_B$ \cite{Petrov2014}, is much smaller than a Bohr magneton $\mu_B$ (approximately the moment of a free electron).  Despite the extremely large electric field, $d_e$ is small enough that $d_e \Eeff$ is always much smaller than $\mu B_z$ for even the smallest magnetic fields that can be reasonably produced in the laboratory.

If the electric and magnetic fields in the precession volume can be accurately and independently reversed, the measured phase differences,
\begin{align}
\phi^E &\equiv \phi(\Eeff,B_z) - \phi(-\Eeff,B_z) = 2d_e\,\Eeff\,\tau/\hbar\\
\phi^B &\equiv \phi(\Eeff,B_z) - \phi(\Eeff,-B_z) = 2\mu\,B_z\,\tau/\hbar,
\end{align}
determine the eEDM,
\begin{equation}
d_e = \frac{\mu\,B_z}{\Eeff}      \frac{\phi^E}{\phi^B},    
\end{equation}
for measured $\mu$ and $B_z$, and a calculated $\Eeff$.  
A discussion of the elaborate fabric of ``reversals'' used to monitor and suppress systematic errors \cite{AcmeEdmDetails2016} is not repeated here because these reversals are not needed to determine the magnetic field requirements for a new measurement.

\subsection{ACME II: Lessons Learned}

For ACME I and ACME II, the precession of ThO was observed over a length $L = 20$ cm, corresponding to a 1 ms precession time.  The five layers of mu-metal shielding used for both measurements were concentric cylinders supported by flat end plates attached to the floor \cite{AcmeEdmDetails2016,SpaunThesis2014,HutzlerThesis2014}. The ThO molecules traveled roughly parallel to the axis of the cylinder.  Each cylindrical layer was split into two half-cylinders that had to be  clamped to each other and attached to the end plates each time they were assembled. The shields were annealed just before the initial construction of ACME I   and then were used for ACME I and II (over nearly 10 years) without being re-annealed.      

Before the ACME II measurements began, the measured residual field inside the spin-precession volume was $50\,\uG$. During the measurement, an internal magnetic field from a cosine theta coil was reversed in direction every 30 to 60 s.  After each field change,  the shields were degaussed with a 2 second procedure  \cite{AcmeEdmDetails2016,LasnerThesis2019}.  By the end of the ACME II measurement, the residual field had risen to $200$--$300 \, \uG$, giving rise to significant systematic errors. About $100 \, \uG$ was attributed to the magnetization of the shields by the field being reversed inside of the shields.  This could be removed with a 10 minute degaussing procedure that was not practical to apply after every magnetic field reversal. Even after this procedure, a residual $200\, \uG$ field remained in the precession volume.  

The source of this residual field is most likely stress-induced magnetism that was introduced as a result of indelicate handling and mounting of the mu-metal components.  The heavy cylindrical shields and the clamping system made it very difficult to assemble and handle the shields without stressing them.  Re-annealing may have removed the lingering residual field.  This was not attempted because of the difficulty, time and cost of transporting large and unwieldy half-cylinders to a large commercial annealing oven, and because the supported end plates could not be easily demounted. 

The ACME II experience suggested several design requirements for an ACME III coil-plus-shields system.  First, it must be feasible to anneal the magnetic shields and even to re-anneal them if this becomes necessary. Second, the shields must be designed so that every step of the assembly and disassembly could be completed without mechanically stressing the shields.  Third, the shields should not support weight other than their own.  These requirements, plus the much large size of ACME III, led to using flat shielding plates rather than cylinders because these can be more readily fabricated and fit into an annealing oven.  Fourth, a robust and effective degaussing system is essential.

\subsection{ACME III Magnetic Field Choice}

ThO molecules from a laser ablation source travel with a range of velocities, $\delta v$, through the length of the precession volume. Measurements and simulations indicate that the mean forward velocity of the molecular beam varied by $0.05\%$ from one ablation-produced pulse of molecules to another \cite{AcmeEdm2018}.  The fluctuations in $v$ are much faster than the reversals of the electric and magnetic field directions.  As a result, there is a distribution of precession times $\delta \tau$, and $\delta \tau/\tau = \delta v/v = \delta \phi/\phi$. Because $\delta \phi^E = \delta \phi^B = \sqrt{2} \delta \phi$ and $d_e \Eeff \ll \mu B_z$ (i.e. $\phi^E \ll \phi^B$), 
\begin{equation}
\delta d_e\approx\frac{\mu B_z}{\Eeff}\frac{\delta \phi^E}{\phi^B} = \frac{1}{\sqrt{2}} \frac{\mu\,B_z}{\Eeff}  \frac{\delta v}{v} 
\label{eq:VelocityUncertainty}
\end{equation}
is the uncertainty in $d_e$ from velocity fluctuations.  

To reach a 40 times lower uncertainty for ACME III compared to ACME II, the velocity-induced uncertainty in Eq.\ \ref{eq:VelocityUncertainty} must be reduced by this factor. Neither $\mu$ nor $\Eeff$ can be changed. The expected $\delta v/v$ will remain the same for a similar ablation source of ThO. The only way to reduce the uncertainty from velocity fluctuations is thus to reduce $B_z$ by a factor of 40.       

The $B_z$ to which this corresponds can be roughly estimated from ACME II.  A factor of 3 ``excess noise'' from velocity fluctuations in ACME II was observed for $B_z = 26$ mG \cite{PandaThesis2019,MeisenhelderThesis2023}.  Reducing the field by twice the excess noise factor to 4 mG, should bring this noise well below other sources.  ACME II used $B_z=0.7$, $1.3$, and $2.6\, \text{mG}$ for measurement, where no excess noise associated with magnetic field was observed.  Reducing a 4 mG field by a factor of 40 gives $100 \, \uG$, the nominal choice for ACME III.  Measurements will, of course, be performed at several different values of $B_z$, to keep excess noise related to velocity uncertainty from limiting the accuracy.

\subsection{ACME III Field Homogeneity and Stability}

ACME II made successful measurements with $B_z = 700 \, \uG$ despite a non-reversing component of $200\,\uG$, only 3.5 times smaller than the applied field.  To be safer, we set our design goal to be a non-reversing field of $10 \, \uG$ -- 10 times smaller than the nominal $B_z=100 \, \uG$.  The magnetic shielding needed for a new measurement must reduce the ambient field to at least this level.  The residual magnetism of the shield should also approach this value.  For a nominally $100 \, \uG$ field, the coil must produce a field that varies by less than 10\% everywhere in the spin precession volume.  

The magnetic field coils to be described can be used to introduce and vary a larger non-reversing magnetic field.  The observed eEDM signal can then be extrapolated to the size of the observed non-reversing magnetic field for uncertainty estimates.

\subsection{ACME III Magnetic Field Gradient}

ACME II reported two systematic shifts associated with magnetic field gradients that were 2 -- 3 times smaller than the dominant sources of systematic error. By both carefully choosing laser detunings and applying compensating magnetic field gradients from coils within the shields to reduce the net gradient in the precession volume to 1~$\uG/\text{cm}$\cite{MeisenhelderThesis2023}, the ACME II team reduced these uncertainties.

For ACME III, we set a target of  1~$\uG/\text{cm}$ for magnetic field gradients produced by the actively-shielded coil and for the residual field within the shields before any compensating gradients are applied.  If needed,  gradient coils located in the same volume as the actively-shielded coil (App.\ \ref{sec:AuxCoils}) can produce compensating gradients.  Additionally, one of the two systematic effects involves a correlation between molecule position and transverse velocity $\partial z/\partial v_z$ due to the divergence of the ThO beam, which will be suppressed by the weak focusing of an electrostatic lens which will be employed in ACME III to steer more ThO molecules into the precession volume. \cite{PandaThesis2019,AcmeLens2022}.

\subsection{ACME III Vacuum Chamber}
\label{sec:VacuumChamber}

\begin{figure}[htbp!]
    \centering
\includegraphics[width=\linewidth]{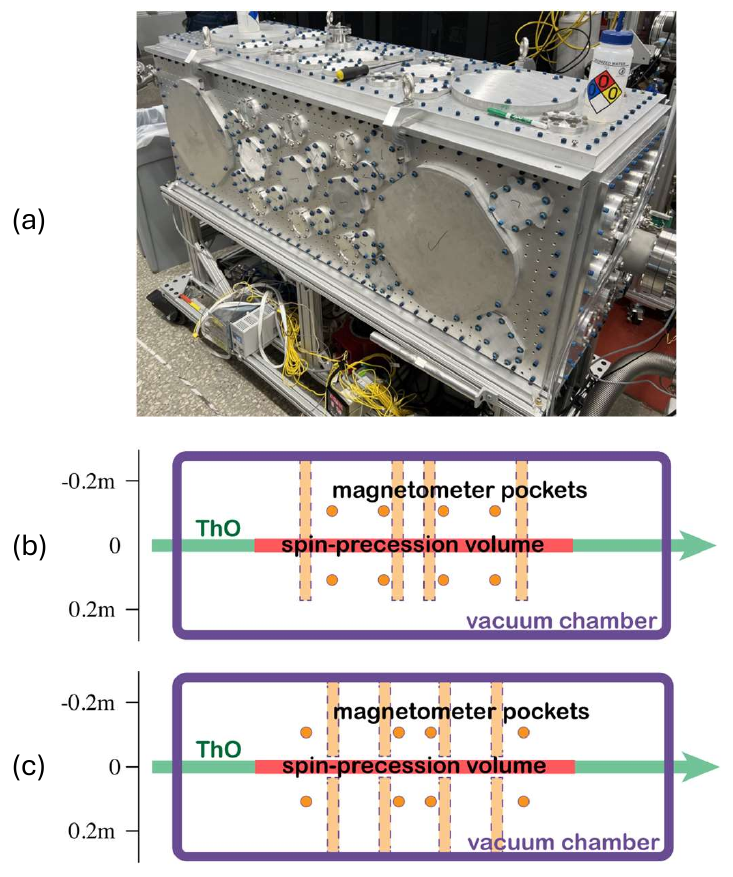} 
         \caption{The ACME III  vacuum chamber (a). Locations of the magnetometer pockets relative to the 1-m-long precession volume are shown in a side view (b)  and a top view (c).}
    \label{fig:Chamber}
\end{figure}

ACME III requires a vacuum chamber large enough to house the 1 m long precession region. This size then constrains the minimum size of the coils. The chamber,  shown in Fig. \ref{fig:Chamber}a, consists of a rectangular welded aluminum frame with dimensions 1.57 m $\times$ 0.58 m $\times$ 0.59 m, which was designed to accommodate the required electric field plates around the precession region. Special care was taken to minimize the magnetism of any materials used in the construction of the vacuum chamber. The vacuum seals were secured by blue-anodized, 2024 aluminum bolts (as brass bolts were detectably magnetic). The bolts threaded into threaded inserts made from titanium in the chamber frame and walls, with PEEK washers to prevent binding of the aluminum screws against the aluminum flanges. 

Access ports on the vacuum chamber have corresponding access holes in the coil and the shields.   
The ThO beam accesses the chamber through an ISO 100 nipple at either end. Six large 14.5 cm diameter window ports provide possible  laser access at both ends of the 1 m long spin precession volume for state preparation and readout. Three smaller 3" (7.6 cm) diameter windows along each side, with 5 similar ports on both the chamber top and bottom, provide additional access options.  Eight 45-degree custom ports couple the signal light from collection optics inside the chamber, through 1" diameter, straight, fused quartz light pipes, to silicon photomultiplier (SiPM) detectors outside the magnetic shields. 

The chamber includes 24 magnetometer ``pockets'' for magnetometers inserted from outside the shields (without breaking vacuum) to monitor the magnetic field as close as possible to the precession volume during measurements. The pockets each have a 2.54 cm inner diameter. Fig. \ref{fig:Chamber}b and \ref{fig:Chamber}c show the locations (5.5 cm from the center axis of the precession volume) in relation to  the spin precession volume.


\section{Actively-Shielded Coil}
\label{sec:Coil}

\subsection{Active-shielding}

 The ACME I--II coil is wound on the surface of a cylinder with a winding density that varies as the cosine of the azimuthal angle \cite{AcmeEdm2014,AcmeEdmDetails2016}.  This produces a magnetic field in the $\zhat$ direction, perpendicular to the direction of travel for the ThO molecules along the $\xhat$ axis of the cylinder. The length of the precession volume was 20 cm.  Sec.\ \ref{sec:ACMEIII} discussed the build up of residual magnetization observed in ACME II, and how a greatly improved ACME III measurement is possible only if this magnetization is dramatically reduced.

The ACME III precession volume (Fig.\ \ref{fig:Overview}) is 5 times longer (1 m $\times$ 4.2 cm $\times$ 4.2 cm). The coil must provide a nominally $100\,\mu$G field perpendicular to its long axis and be spatially uniform to the 10\% level. To make the necessarily larger ACME III system as compact as possible, a rectangular coil assembly is only about 10 cm outside the rectangular vacuum chamber, and is only about 10 cm away from inner mu-metal shield that encloses it. The close proximity of the coil and closest shield raises the concern that the fringe field of the coil could produce residual magnetism in the shields -- as observed in ACME II.  
 
A key advance for reducing this shield magnetism is an  ``actively-shielding'' coil design, with windings designed so that the net field from all the windings largely cancels outside the outer windings. The coil's fringing fields at the location of the nearby mu-metal shields are thereby made small enough to minimize magetizing the shield, and there is  no need to degauss the shields after each of the many magnetic field reversals as was required in ACME II.    

The idea to use active shielding was triggered by the impressively low fringing field of an actively-shielded, high-field solenoid that some of us designed \cite{Gabrielse1988, Gabrielse1991}. The basic idea of active shielding is easy to visualize for solenoids and MRI magnets \cite{ActiveShieldingMRI1,ActiveShieldingMRI2}. Current is sent in opposite directions through two sets of coaxial solenoid layers. The oppositely directed fields cancel significantly outside the solenoid, while only slightly reducing the magnetic field produced inside. 

\begin{figure}[htbp!]
\includegraphics[width=0.6\columnwidth]{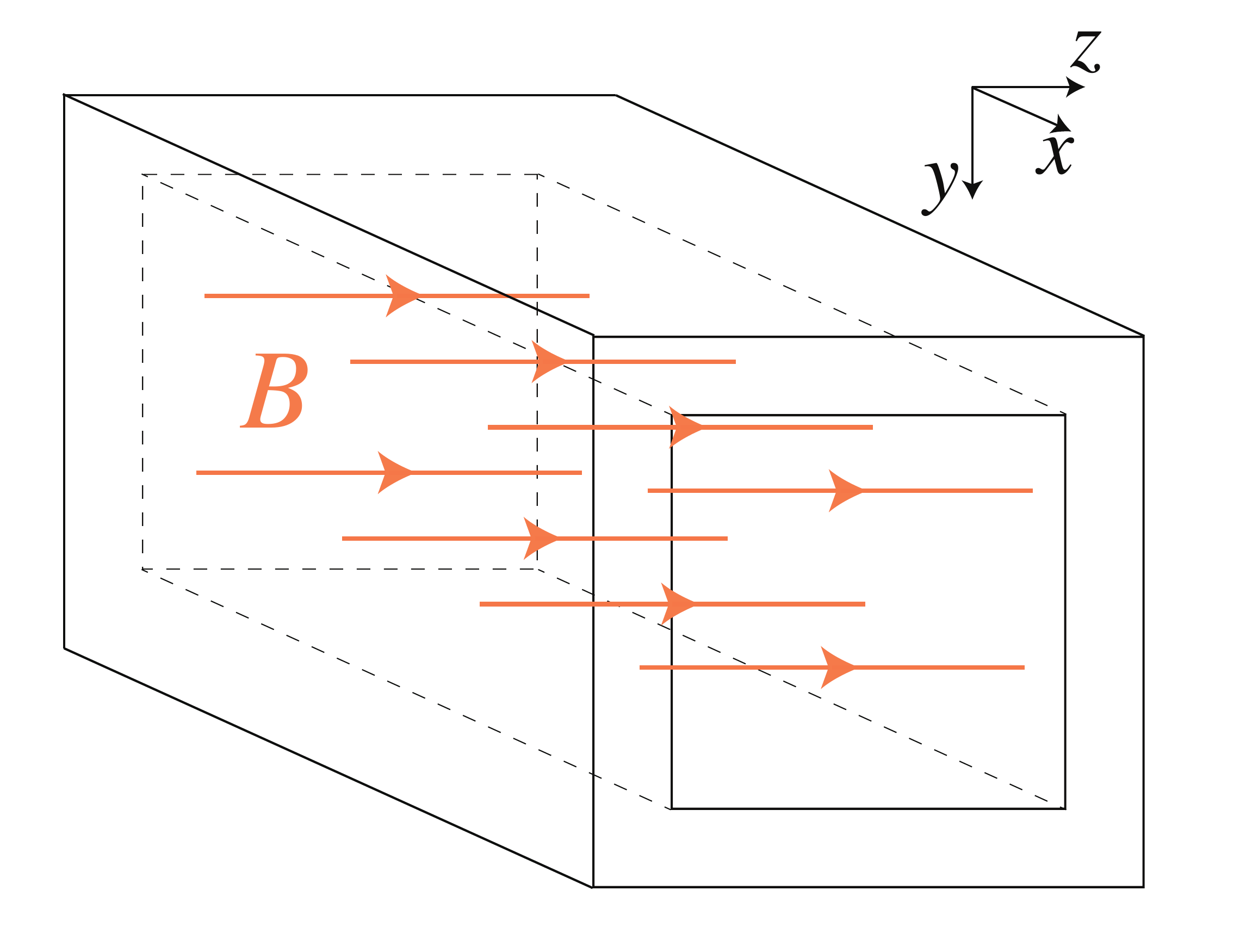}
\caption{Surface currents on the boundaries of two nested 3D rectangular prisms generate $\mathbf{B}=B_0 \, \hat{z}$ inside the inner prism, and ideally no field outside the outer prism.}
\label{fig:NestedRectangles}
\end{figure}

Active shielding for an ACME eEDM measurement requires a coil geometry that is very different from a solenoid because the field must be perpendicular rather than parallel to the coil's long axis. The oppositely directed currents needed for active shielding in this case flow along the inner and outer surfaces of nested rectangular solids (Fig.\ \ref{fig:NestedRectangles}).  
The magnetic fields that we wish to produce inside the inner surface, $\mathbf{B}=B_0 \, \hat{z}$, and outside the outer surface, $\mathbf{B} = 0$, determine the surface currents that must flow.

We utilized a novel mathematical approach for designing self-shielding coils \cite{crawfordshinbfield, crawfordmartinbfield}. Two familiar boundary conditions, from Gauss' Law and from Ampere's Law, relate the magnetic fields on either side of a boundary surface to the surface currents that must flow.  The volume inside each of our boundary surfaces contains the precession volume. The field in this volume, just inside the boundary surface, is $\Bin$.  The field just outside the boundary surface is $\Bout$. The unit vector $\nout$ on each boundary face is normal to the surface.  On each face it points outward and away from the inner shielded volume that includes the origin. 
From Gauss' Law, 
\begin{equation}
\nout \cdot (\Bin - \Bout) = 0.
\label{eq:GaussBoundary}
\end{equation}
The normal component of the magnetic field is thus always continuous across a boundary surface. 
From Ampere's Law, a surface current density (current per transverse length), 
\begin{equation}
\K = (\Bin - \Bout ) \times \nout,
\label{eq:AmpereBoundary}
\end{equation}
must flow along the boundary surface for there to be a discontinuity in the  tangential magnetic field.

\subsection{Magnetostatics Boundary Value Problem}

\begin{figure}[htbp!]
\includegraphics[width=\the\columnwidth]{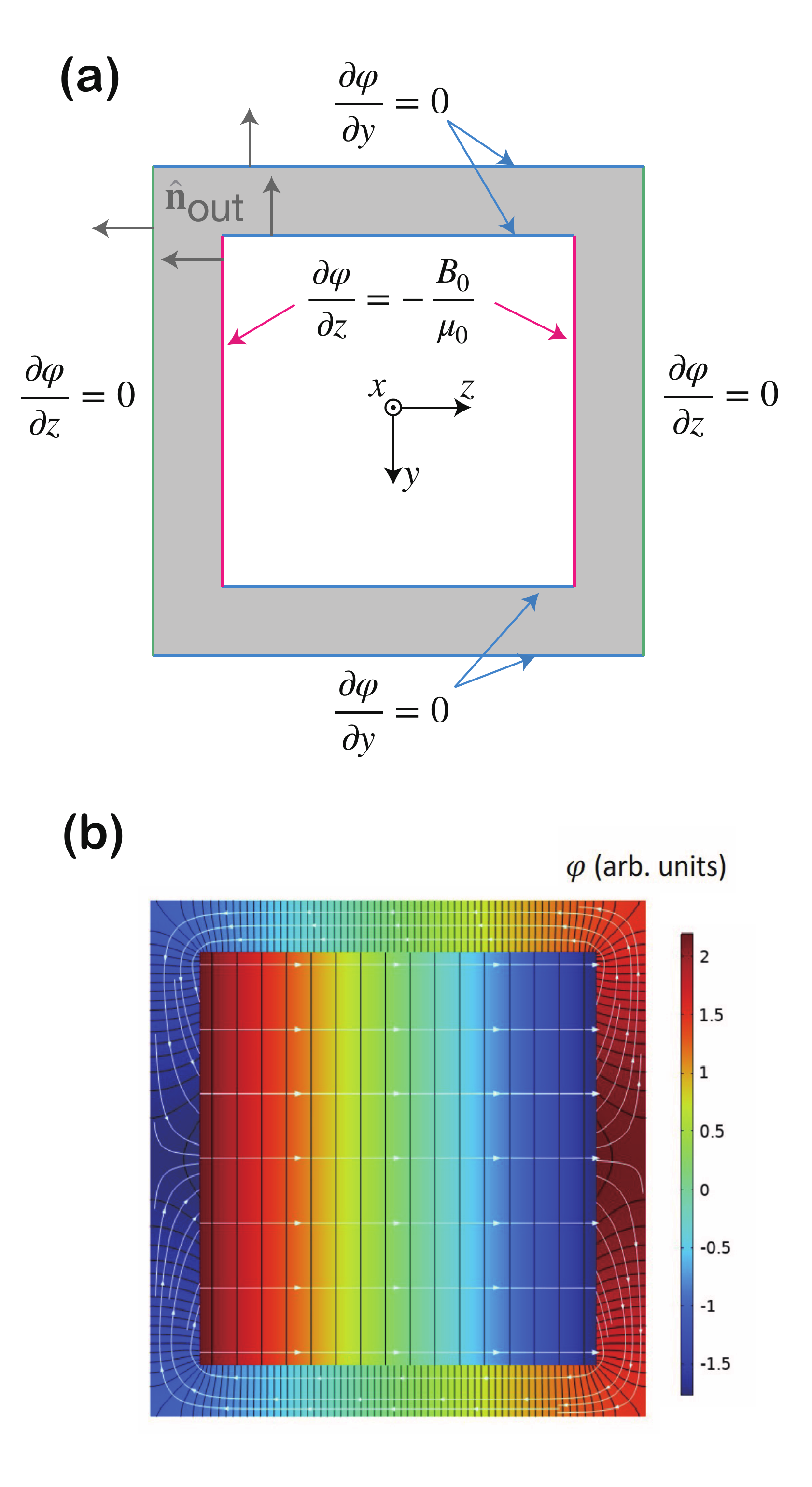}
\caption{(a) Boundary conditions for $\varphi$ in the yz plane. (b) Equipotential lines of the solution $\varphi$ of  the Laplace equation between the inner and outer coil boundaries, with the assumed $\varphi$ outside the outer boundary and inside the inner boundary.  The color represents the magnitude of  $\varphi$.} 
\label{fig:ScalarPotential}
\end{figure}

Current is only allowed to flow on the inner and outer boundaries of the nested 3D prisms in Fig.\ \ref{fig:NestedRectangles}. The magnetic field in the volume between the nested rectangles (gray in Fig.\ \ref{fig:ScalarPotential}a) can be written as the gradient of a magnetic scalar potential,  
\begin{equation}
    \mathbf{B}=- \grad \varphi.   \label{eq:magneticPotentialDefinition}
    \end{equation}
Because no current flows within this volume, Ampere’s law gives $\nabla\times\mathbf B=0$, while Gauss' Law gives $\nabla\cdot\mathbf B=0$, ensuring that $\varphi$ is a solution to Laplace's equation, $\nabla^2\varphi=0$.  Any solution that satisfies appropriate boundary conditions on the surrounding boundaries is the unique solution  
\cite{jackson1998classical}.  

There are Neumann boundary conditions on the normal derivative of $\varphi$ for each of the faces that bound the volume of interest. Faces that are part of or touch the outer 3D rectangle all have $\Bout=0$, which means that the normal derivative of $\varphi$ at these faces vanishes.  The two faces on the inner 3D rectangle that are parallel to the $xz$ plane have $\Bin=B_0 \zhat$ and $\partial \varphi / \partial y = 0$. The two faces on the inner 3D rectangle that are parallel to the $xy$ plane have $\Bin=B_0 \zhat $ and $\partial \varphi / \partial z = -B_0$. The boundary conditions for most of the boundary faces, viewed end on, are represented in Fig.\ \ref{fig:ScalarPotential}.

The scalar potential $\varphi$ is computed everywhere in the volume of interest using the COMSOL \cite{COMSOL} finite element analysis computer program, for the listed boundary conditions. The solution in the central $x=0$ plane is represented in Fig. \ref{fig:ScalarPotential}b.

\subsection{Current Distribution and Wiring Pattern}

What remains is to apply the Ampere's law boundary condition in Eq.\ \ref{eq:AmpereBoundary} to each boundary face, to determine the surface current density $\K$ that must flow on that face. On all boundary faces that are on or touch the outer 3D rectangular prism, 
\begin{equation}
\K = -\grad \varphi \times \nout,
\label{eq:OuterCurrent}
\end{equation}
where $\grad \varphi$ and $\nout$ are evaluated for each face. Because $\K$ is perpendicular to $\grad \varphi$, which is  perpendicular to equipotentials of $\varphi$, the surface currents in the flat faces are always directed along equipotentials of $\varphi$ in each face.  (As stressed in \cite{CrawfordRSI2021}, this is also a general feature for surface currents flowing on boundaries that are not flat.) 

On the boundary faces on the inner 3D prism, there is a surface current density
\begin{equation}
\K = B_0 \zhat \times \nout + \grad \varphi \times \nout, 
\label{eq:InnerCurrent}
\end{equation}
where $\grad \varphi$ and $\nout$ are evaluated on each face.  
The first term is a uniform $\K$ that tangentially encircles $B_0 \zhat$. The second term, like the current densities on the outer boundary faces described by  Eq.\ \ref{eq:OuterCurrent}, is directed along equipotentials of $\varphi$ on the faces of the inner 3D rectangle.  

A recent paper \cite{CrawfordRSI2021} emphasizes that a magnetostatics boundary value problem can be solved to find a scalar potential and surface currents that must flow on boundaries of arbitrary shape -- not just on the simple flat faces of our nested 3D rectangles.  That surface currents flow in the direction of the equipotentials of the scalar potential is a general and useful feature.  

\begin{figure}[htbp!]
\includegraphics[width=\the\columnwidth]{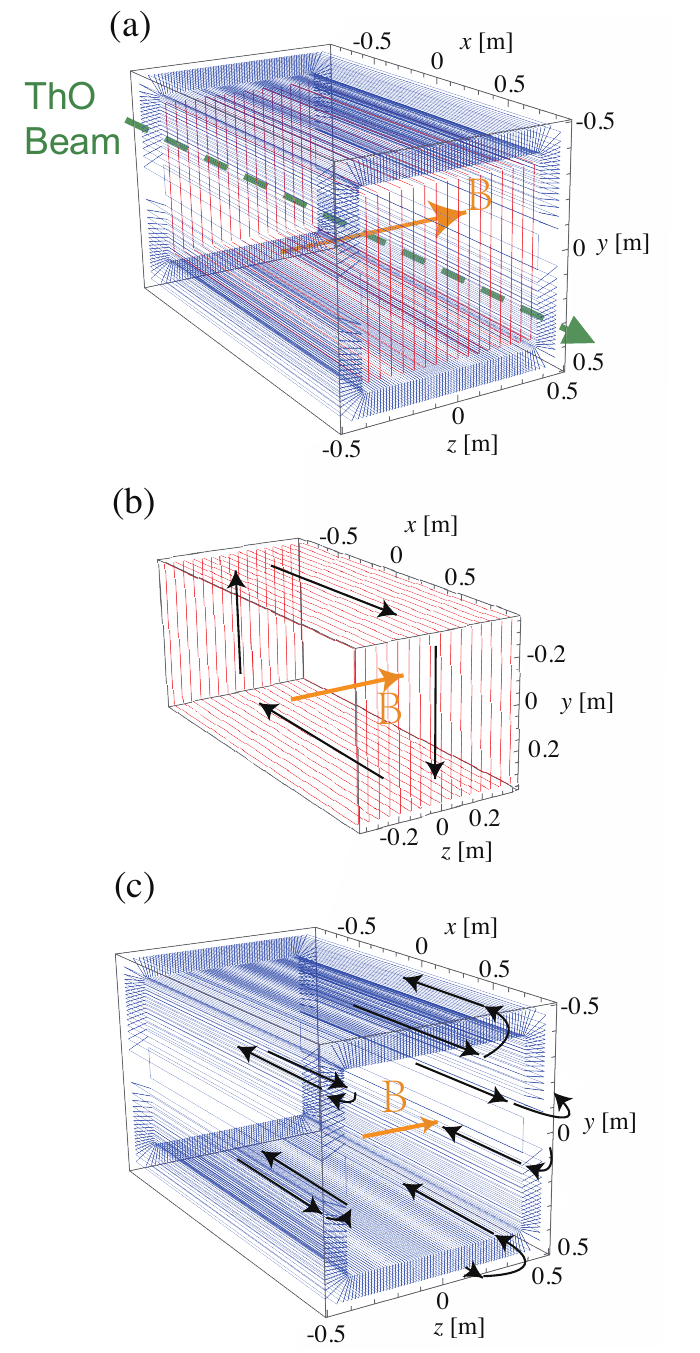}
\caption{The ACME III $B_z$ magnetic field coil design obtained after finite element analysis for the combined coils (a), the inner coil (b), and the outer coil (c). Black arrows indicate direction of current flow. Minor wire rerouting paths  around access openings through the coils are not shown.}
\label{fig:CoilWindings}
\end{figure}

\subsection{Coil Windings Approximate Surface Currents}

We approximate the desired surface currents in Eqs.\ \ref{eq:OuterCurrent} -- \ref{eq:InnerCurrent} using discrete coil windings that follow equipotentials of $\varphi$ on the flat boundary faces, and are spaced so that the same current can be used in all of the windings.  Fig.\ \ref{fig:CoilWindings} represents the 198 wire loops used for the two-layer coil (Fig.\ \ref{fig:CoilWindings}b), and the 16 windings in the single-layer coil (Fig.\ \ref{fig:CoilWindings}c). Current in the 16 wire loops of a single layer coil represented in red in Fig.\ \ref{fig:CoilWindings}b approximates the first term in the current density of Eq.\ \ref{eq:InnerCurrent}.  This same current is sent through the 198 current loop windings of the blue coil in Fig.\ \ref{fig:CoilWindings}c.  The inner and outer windings approximate the current densities described in the second term of Eq.\ \ref{eq:InnerCurrent} and Eq.\ \ref{eq:OuterCurrent}, respectively. The equipotentials connecting the layers of wires in the two-layer blue coil of Fig.~\ref{fig:CoilWindings}c are nearly linear segments, and are approximated as such in the coil construction.

\begin{figure}[htbp!]
\includegraphics[width=1\columnwidth]{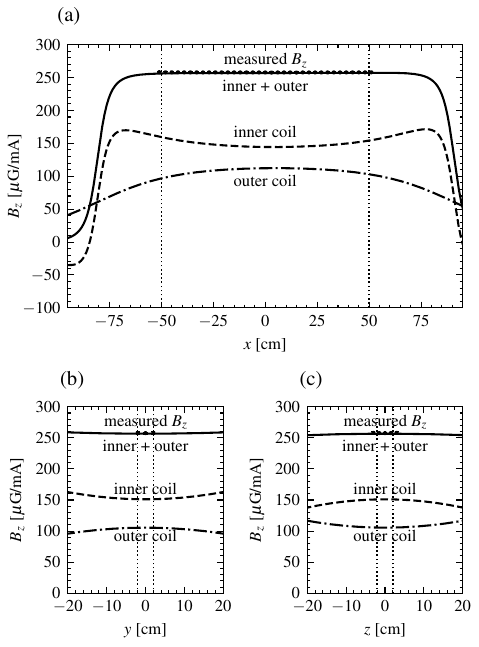}
\caption{Calculated $B_z$ along the $x$ axis (a), the $y$ axis (b) and the  $z$ axis (c) for the coil windings in Fig.\ \ref{fig:CoilWindings}.  The contributions of the outer two-layer coil and the inner single-layer coil are shown, and solid curves are their sums. The vertical dotted lines mark the edges of the spin-precession volume.  The measured values and uncertainties in (a) agree so well that they are difficult to distinguish.}
\label{fig:CoilContributions}
\end{figure}

As a check, the Biot-Savart law (implemented in the Radia computer package \cite{Chubar1998}) is used to calculate the magnetic field produced by the chosen coil windings in the precession volume. This allows us to confirm that the number of wires is large enough to achieve both the desired magnetic field homogeneity and a low calculated fringe field at the location of the inner layer of magnetic shielding. We use the ``as built'' wire paths that go around the openings in the coil that give  access to the precession volume from outside of the coil system.  Fig.\ \ref{fig:CoilContributions} shows how the fields of the two-layer and the single layer coils combine to make a spatially uniform $B_z$. 

The calculated magnetic field in the precession volume when no magnetic shielding is present is $257 \pm 1\,\uG/\text{mA}$.  A readily-produced 389 $\mu$A current thus produces the nominal $100\, \mu$G field.  (We shall see in Sec.\ \ref{sec:Performance} that the measured $257 \pm 2\,\uG/\text{mA}$ is in good agreement.)  The calculated homogeneity over the 1 m $\times$ 4.2 cm $\times$ 4.2 cm precession volume is 0.33\%. This is $0.33\,\uG$ for the nominal $B_z=100 \, \uG$, significantly better than the  $10\mu\text{G}$ design goal. 

The active shielding design and the lower anticipated $B_z$ reduce the fringe field that can magnetize the nearest shield by about a factor of $150$ compared to ACME II. Let $|B_{s}|$ be the largest magnitude of the fringing field calculated on the surface where the nearest mu-metal layer is to be located. The ratio $|B_{s}|/|B_z|$ is an approximate but useful figure of merit for the active shielding. The  size of this 3D rectangle is  $1.86 \,\text{m} \times 1.17 \,\text{m}\times 1.17 \,\text{m}$.  Our actively-shielded coil produces the small ratio, $|B_{s}|/|B_z| = 5.3\%$.  This is only $5.3 \,\uG$ for the nominal  $100 \, \uG$ planned for ACME III.  For ACME II,  the calculated $|B_{s}|/|B_z| = 78\%$ is 15 times larger.   If we produce $B_z$ using only the  inner red coil in Fig.\ \ref{fig:CoilWindings}b (not actively shielded by itself), the largest calculated   $|B_{s}|/|B_z| = 66\%$ is 13 times larger (on the faces parallel to the xy plane).  
The $B_{s}$ will be smaller by an additional factor of 10, compared to ACME II, because we anticipate using a $B_z$ that is smaller by approximately this factor.

\subsection{Implementation}

\begin{figure}[htbp!]
\includegraphics[width=\the\columnwidth]{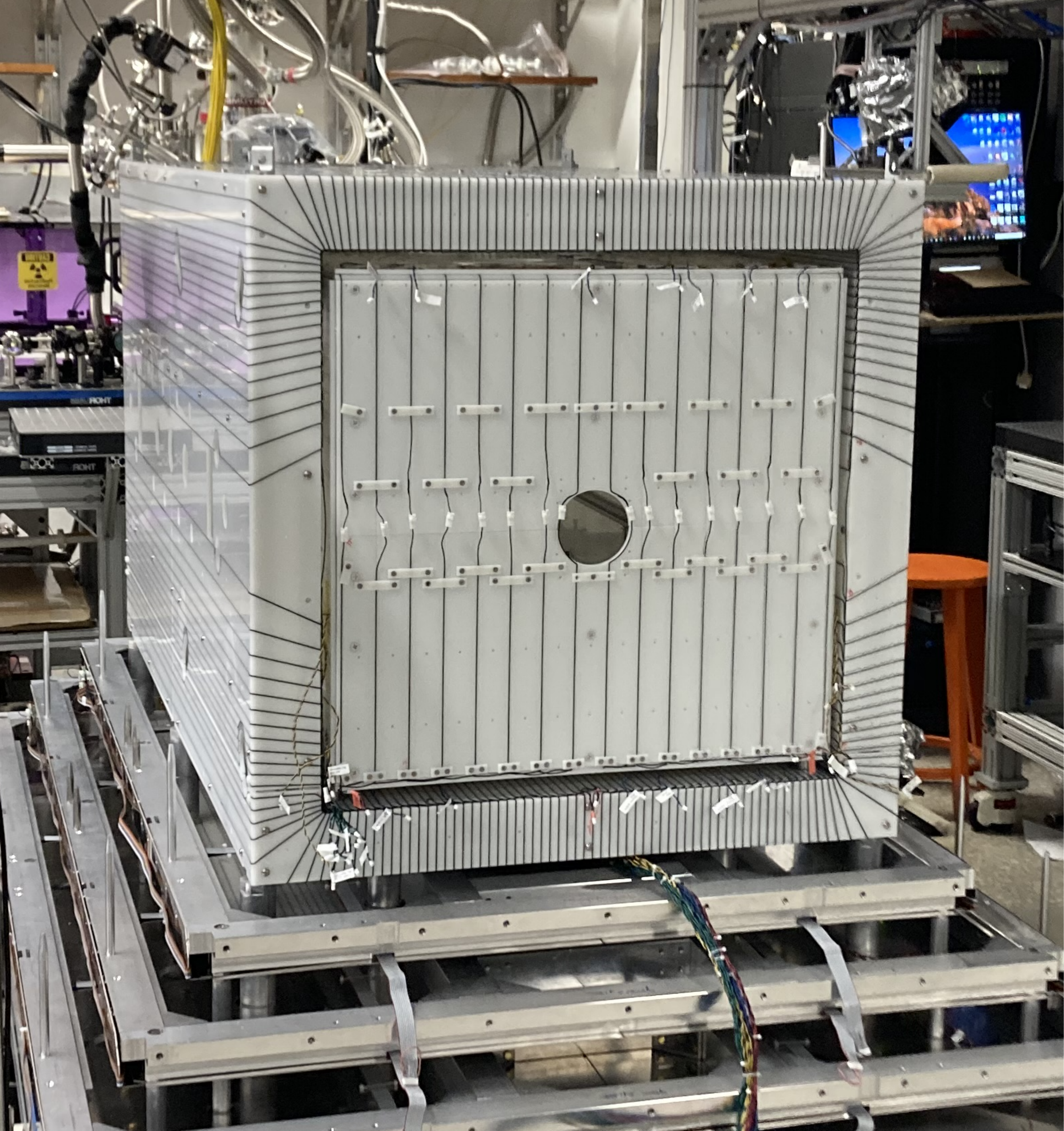}
\caption{The actively-shielded two-layer coil has outer dimensions $1.72\,\text{m}\times 0.975\,\text{m}\times 0.975\,\text{m}$. }
\label{fig:CoilPicture}
\end{figure}

Fig.~\ref{fig:CoilPicture} shows the actively-shielded coil.  Insulated solid copper wire (18 AWG) is pushed gently into grooves machined into 1/4 inch thick high-density polyethylene sheets, supported by 1 inch 8020 aluminium frames.  Bolts holding this assembly together are made of titanium to avoid having their residual magnetism contribute to the magnetic field on the $\mu$G level. (Brass bolts were found to be too magnetic at the $\uG$ level.)   The coil is fully demountable so that the vacuum chamber can be inserted and accessed inside. This requires disconnecting 81 wires (34 for the active shielding coils and the rest for the gradient coils discussed in Sec.\ \ref{sec:AuxCoils}). Socket and pin connectors (of types SPHD-001T-P0.5 and SPAL-001T-P0.5 pins and PALR-02VF and PAP-02V-S housings) facilitate this and have served robustly.  

The wire locations were chosen such that one power supply (Stanford Research Systems Inc. CS-580)  sends the same current through all of the wires in series.  The measured wire resistance is 18 $\Omega$ for the outer main coil and 3.2 $\Omega$ for the inner main coil. As will be discussed in App.\ \ref{sec:AuxCoils}, to allow us to make a gradient in $B_z$ that varies in $z$, the actively-shielded coil is wired so that we can send different (or the same) currents through the turns that are to either side of the center along the $z$ axis. Each of the two currents is separately measured from the voltage drop across precision 1 $\Omega$ shunt resistors.


\section{Magnetic Shields}
\label{sec:MagneticShields}

\subsection{Shielding Requirements}
\label{sec:ShieldingImplementation} 

A three layer magnetic shield (Fig.\ \ref{fig:Overview}) is added to keep the ambient magnetic field in the lab from changing the magnetic field in the precession volume by more than the required $10\,\uG$ in space and time.  For ACME, it is crucial that both the ambient field and the shields' remanent magnetization do not create a field that fails to reverse when the coil current changes direction.

\begin{figure}
    \centering
\includegraphics[width=\columnwidth]{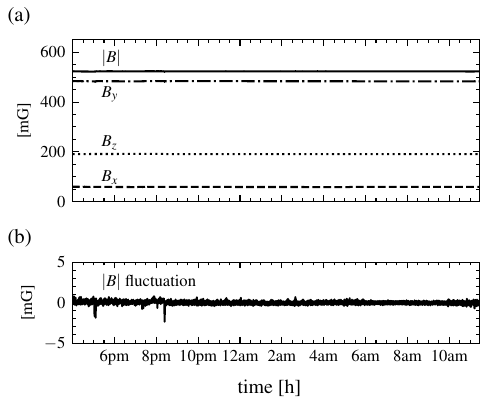}
    \caption{Ambient field 2 m above the ACME III apparatus (a) has fluctuations (b) that are much smaller than the average field. }
    \label{fig:AmbientB}
\end{figure}

One example of the ambient magnetic field, shown in Fig.\ \ref{fig:AmbientB}, is measured 2 m above the coil-plus-shields system. In this example, the measured average is $522 \pm 2$ mG. This is primarily the earth's field, but fields averaging between 0.4 and 0.6 G are measured at nearby locations in the lab, depending upon the proximity of magnetized equipment and apparatus. The ambient field must thus be smaller everywhere in the long precession volume by about a factor of $10^5$. A smaller shielding factor is required for typical ambient field fluctuations in time given that these are typically so much smaller than the average. 

A 4 s degaussing procedure that achieves the $10^5$ reduction is described in Sec.\ \ref{sec:Degaussing}.  Past electric dipole moment measurements, including ACME II, typically degaussed mu-metal shields every time the internal magnetic field was changed \cite{AcmeEdm2014,neutronEDMMagneticShield2017ORNL,neutronEDMShield2014PSI,neutronEDMShield2015TUM,neutronEDMTUMMagneticShield2014}.   The ACME III coil-plus-shields system was designed with the goal of greatly reducing how often degaussing is required, and measurements reported in Sec.\ \ref{sec:Performance} show the feasibility of this goal.

A special challenge in the Northwestern University lab in which our system is located is an unshielded 70 kG (7 T) solenoid that is 6 m away from our precession volume.  When the solenoid current is ramped up and down, it produces a fringe field at the shield location between $-260$ mG and $260$ mG.  This addition is comparable to the ambient field before the solenoid is energized.  We shall show in Sec.\ \ref{sec:Performance} that changes in even this large field are reduced by $1.8 \times 10^4$ in the precession volume.  After the shields are degaussed while the solenoid is operated at any fixed field, the solenoid field is no longer observed.

\subsection{Implementation}
\label{sec:ShieldImplementation}

To minimize the cost and weight of the mu-metal, we designed and produced the coil-plus-shields system to be as compact as was possible.  The nominal separation between the vacuum chamber, actively-shielded coil system, and the mu-metal shields was only 10 cm (Fig.\ \ref{fig:Overview}).

To realize the highest permeability and the lowest remanent field, mu-metal shields must be annealed -- by heating them to 1121 $^\circ$C in a high purity hydrogen atmosphere and carefully controlling their cool down rate.  After this, they must be transported and handled very carefully to avoid stress- or shock-induced magnetism and loss of permeability that would require the shield to be re-annealed  \cite{Preece1971}.  

To this end, our shields were designed to be fully demountable.  They were constructed mostly of the largest flat plates that could be fit into the largest commercial annealing oven to which we could purchase access (Exotic Metal Treating Inc.). Our largest plate is 2.18 m $\times$ 0.75 m $\times$ 1.6 mm and weighs 19 kg (42 lbs).   Each plate can thus be safely handled by two people for transportation, and for assembly and disassembly without adding undue stress or shock.  To ensure proper handling, we transported the 1.4 tons (3000 lbs) of mu-metal ourselves in a small rental truck that we drove 200 miles to the annealing facility. Mu-metal sheets were stored in a wooden crate and were insulated from the crate walls with thick foam. They were separated from each other with packing paper. 

Before annealing, the rectangular plates were cut to size with a water jet. All cut edges were deburred by hand.  Some of the 4185 holes and slots needed for mounting and for providing line-of-sight access through all three shielding layers to the vacuum chamber (e.g.\ for lasers, light pipes and magnetometers) are visible in Fig.\ \ref{fig:OneFaceDegaussing}. Also visible are the access slots (up to 6.3 cm wide $\times$ 14 cm high) and circular holes up to a 9 cm diameter.  These dimensions were chosen, based upon finite element analysis simulations, so that the ambient field flux leaking through the openings would not distort the magnetic field in the precession region at the $\uG$ level.  

To ensure a continuous magnetic flux path all around each shield layer, the flat metal plates all overlap by 4 to 8 cm. A separation of $10.0\,\text{cm}$ is designed between each layer in all directions.  Every corner and edge of the shields is covered with custom-fabricated mu-metal bridge pieces bent $90^\circ$ at nearly the specified $1.6\,\text{mm}$ radius, with a similar overlap with the flat sheets.

Magnetic Shields Inc.\  supplied the 1.6 mm thick sheets of "CoNetic AA Perfection Annealed alloy" from which we cut the plates \cite{CoNeticAA}. (Note:  both MuMetal and CoNetic are brand names and the generic term ``mu-metal'' is often used for both.)  The saturation specification is listed at 7.5 kG (0.75 T).  The magnetic shielding properties of this same material have been independently measured \cite{Arpaia2021} and after annealing, an initial permeability of $50,000$ has been measured. After an ideal degaussing process, the directions of the magnetic domains reach thermodynamic equilibrium with the external field \cite{MagneticShieldTextbook_2013}.  The resulting anhysteretic magnetization curve (B versus H), provided as a specification, has $\mu_r = 350,000$ for a field H=0.  

\begin{figure}[htbp]
  \centering
 \includegraphics[width=\columnwidth]{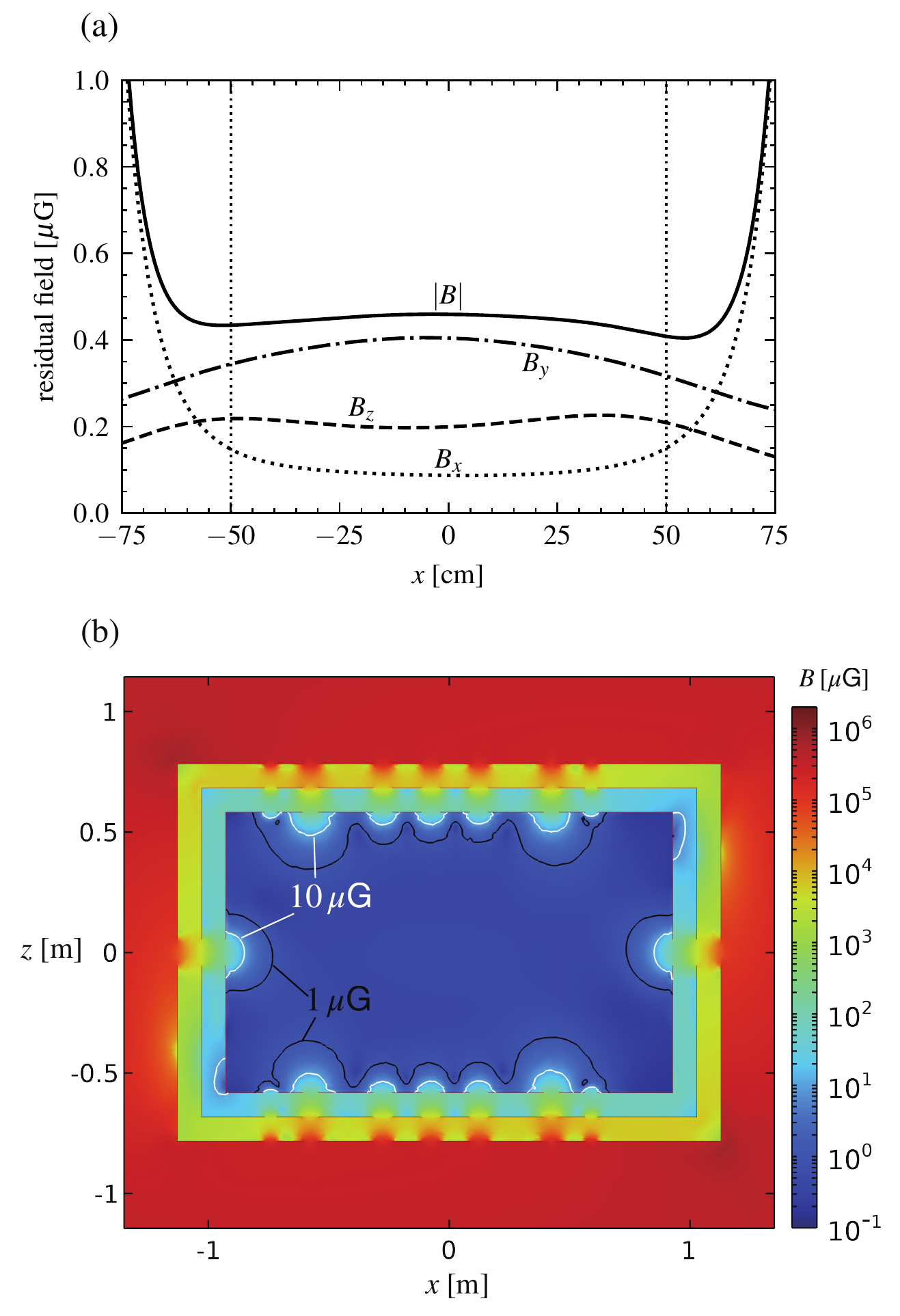}
  \caption{The calculated field components along the central $x$ axis (a), and the calculated field magnitude in the $xy$ plane (b), for the 0.5 G ambient field of Fig.\ \ref{fig:AmbientB}a,}
    \label{fig:CalculatedField}
\end{figure}

The shielding provided by the rectangular shields can only be calculated numerically.  
We performed a 3D finite analysis calculation (using COMSOL \cite{COMSOL}) using the anhysteretic curve provided by the mu-metal supplier.  The result was that a three-layer rectangular shield should be able to achieve the targeted $10^5$ suppression of the ambient field. Fig.\ \ref{fig:CalculatedField}a shows the magnetic field components and magnitude calculated along the central $x$ axis when the external ambient field in Fig.\ \ref{fig:AmbientB}a is applied outside the magnetic shields.  Fig.\ \ref{fig:CalculatedField}b represents the field magnitude everywhere in the $xy$ plane.  

\begin{figure}
    \centering    \includegraphics[width=\columnwidth]{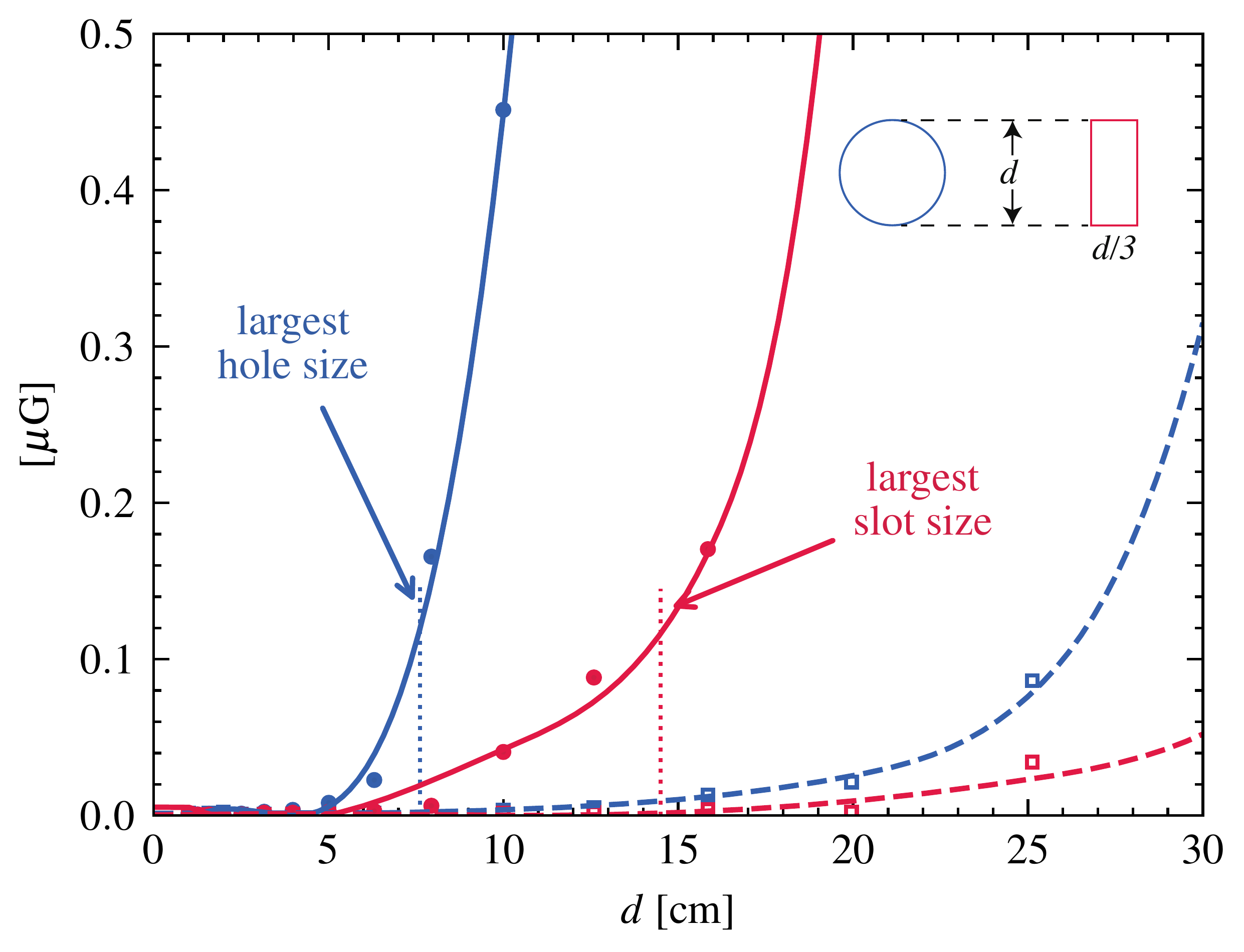}
    \caption{For a 0.5 G ambient field, the calculated field at the intersection of the $x$ axis and the axis of a single set of access openings in the 3 shielding layers increases with the diameter $d$ of a circular opening (blue), and for a slot of height $d$ and a width $d/3$ (red). It is larger for an ambient field that is perpendicular to the plane of the opening (solid curves), than for an ambient field parallel to this plane (dashed curves).}
    \label{fig:Holes}
\end{figure}

To evaluate the effect of the size of access holes in the shields, we calculated the effect of a single hole or slot in the three layers of shields for an ambient field of 0.5 G applied both parallel and perpendicular to the plane of the hole. Fig.\ref{fig:Holes} shows the field magnitude on the central $x$ axis at the point centered on the opening as a function of the diameter $d$ of a circular opening, and for a slot of height $d$ and a width $d/3$.

\begin{figure}[htbp]
  \centering
  \includegraphics[width=\columnwidth]{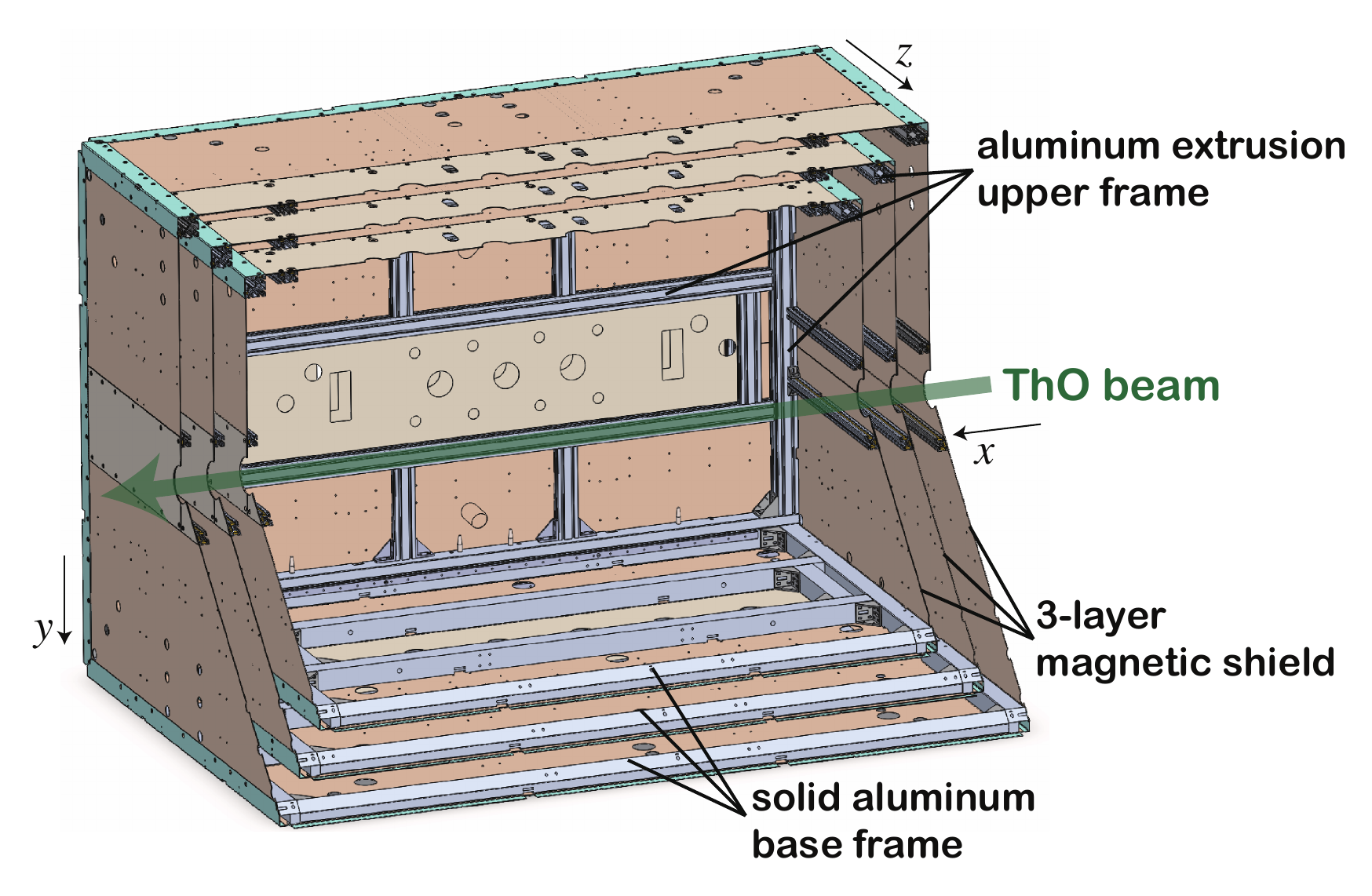}
  \caption{Cross-section of half the 3-shield system, to scale.  The inner layer is $1.857 \,\text{m}\times 1.167 \,\text{m} \times 1.167 \,\text{m}$, with its longer dimension oriented to align with the molecular beam pathway of the ACME experiment. }
    \label{fig:ShieldsCutaway}
\end{figure}

The plates were gently attached to aluminum supports to form them into rectangular shielding layers (e.g.\ Fig.\ \ref{fig:ShieldsCutaway}) without stressing or shocking the plates. Each of the 2028 3/8 in mounting bolts is passed through a hole in the mu-metal that is large enough to minimize contact between the bolt and mu-metal (which would stress the mu-metal).  Each bolt is tightened against a washer by hand to a torque of $3.4\,\text{N}\cdot\text{m}$, and threaded into custom aluminum plate-nuts inserted into an aluminum frame. Titanium bolts and washers were used for the inner layer (because brass bolts were too magnetic), but brass bolts and washers were used for the outer two layers. 
Fig.\ \ref{fig:ShieldsCutaway} is a cutaway diagram showing one half of the top and sides of the three layers of mu-metal (brown and green). 

\begin{figure}[htbp]
  \centering
  \includegraphics[width=\linewidth]{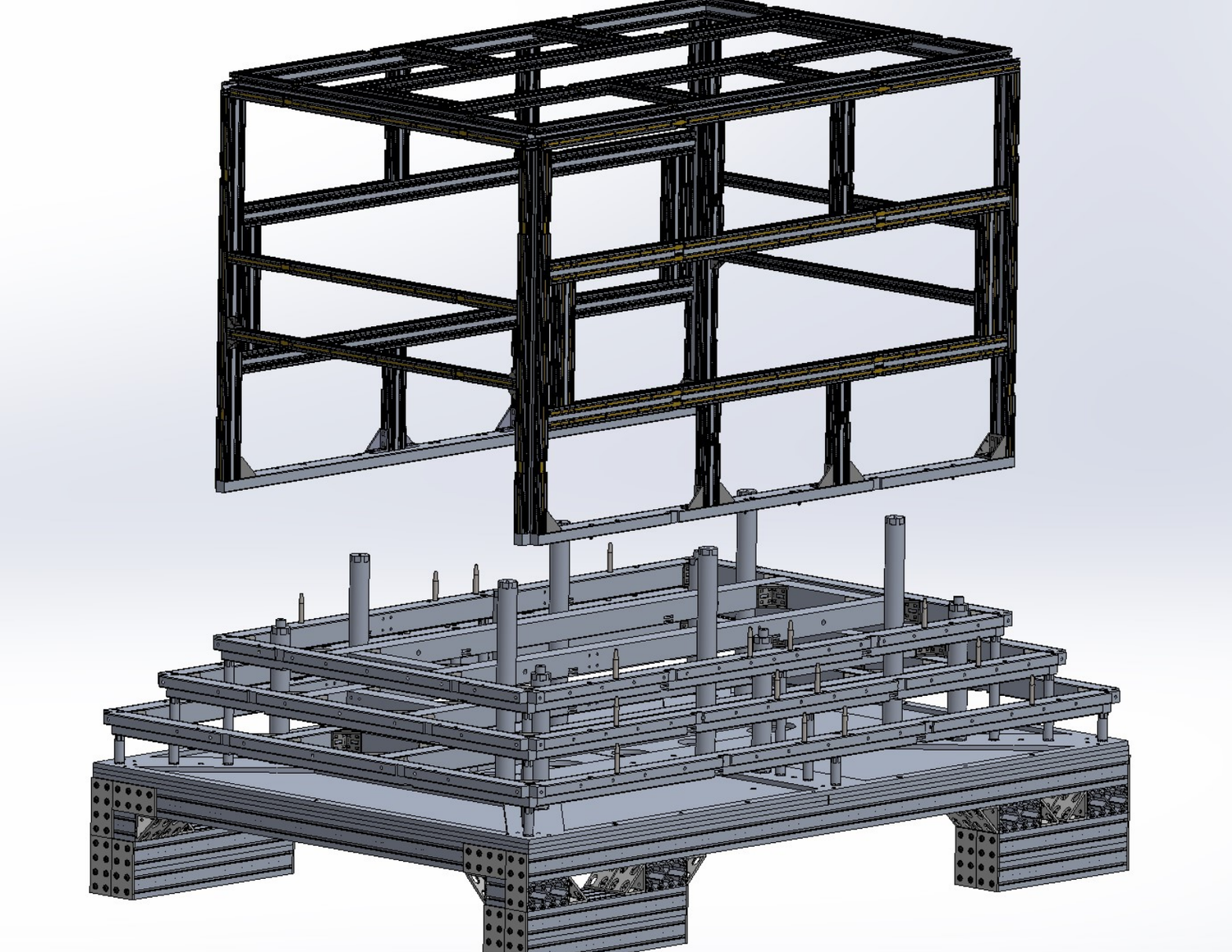}
  \caption{The frame for the top and side layers of magnetic shielding  can be lifted off as a unit.}
    \label{fig:UDesign}
\end{figure}

The top and sides of each shielding layer can be removed together as a unit to make it possible to maintain, update and remove the coil, vacuum chamber, and the apparatus inside it. Fig.\ \ref{fig:UDesign} illustrates how the frame is lifted, although the shielding plates that remain attached during the lift are not shown. Once it is lifted, what remains is a "table" that includes the bottom layers of each of the three shields (not shown in Fig. \ref{fig:UDesign}). 
 
 The table is designed to support 1225 kg (2700 lbs) of shields, a 189 kg (416 lbs) active-shielding coil, and a 450 kg (990 lbs) chamber.  Each of the shield layers is supported by 1'' diameter aluminum rods that connect the table to the bottom frame of the layer.  The coil and chamber are supported by longer, 2'' diameter aluminum rods.  The support rods for each layer pass through clearance holes in the layers outside it.  The holes are sized to make sure there can be no contact.  A system of alignment rods position the shields in the proper location without stressing the mu-metal when these heavy assemblies are lowered into position. 

The rectangular shape of the shields is similar to what is used for magnetic shielded rooms (MSRs). Examples include a 5-layer MSR at TUM for a neutron EDM experiment\cite{neutronEDMShield2015TUM}, a 2-layer shield with active field cancellation for magnetoencephalography (MEG) \cite{Holmes2022}, and the 8-layer BMSR-2 at PTB Berlin for ultra-low magnetic field metrology \cite{inproceedings_BMSR} and other biomedical sensing applications. MSRs typically cannot have sections lifted off or be fully demounted, relying instead upon an entrance door. Shielded rooms are often much larger than the experiments carried out inside them, particularly if the experiments inside produce magnetic fields.

\subsection{Degaussing}
\label{sec:Degaussing}

After exposure to an external magnetic field, high-permeability ferromagnetic alloys, such as the Co-Netic AA we use, retain a degree of remanent magnetization after this field is removed. Degaussing is a procedure that randomizes the magnetic domains in the mu-metal to decrease the remanent field\cite{thiel_demagnetization_2007}. 

For ACME II (Sec.\ \ref{sec:ACMEIII}), a 2 s degaussing procedure was applied every time the magnetic field was changed\cite{LasnerThesis2019}.  Degaussing thus consumed a bit more than 6\% of the possible measurement time.  Nonetheless, a $300 \, \uG$ remnant field built up during repeated reversals of the internal applied field -- much larger than the $10 \,\uG$ required for ACME III.  Even after much more extensive degaussing, a $200\, \uG$ non-reversing field remained.

\begin{figure}[htbp]
  \includegraphics[width=\linewidth]{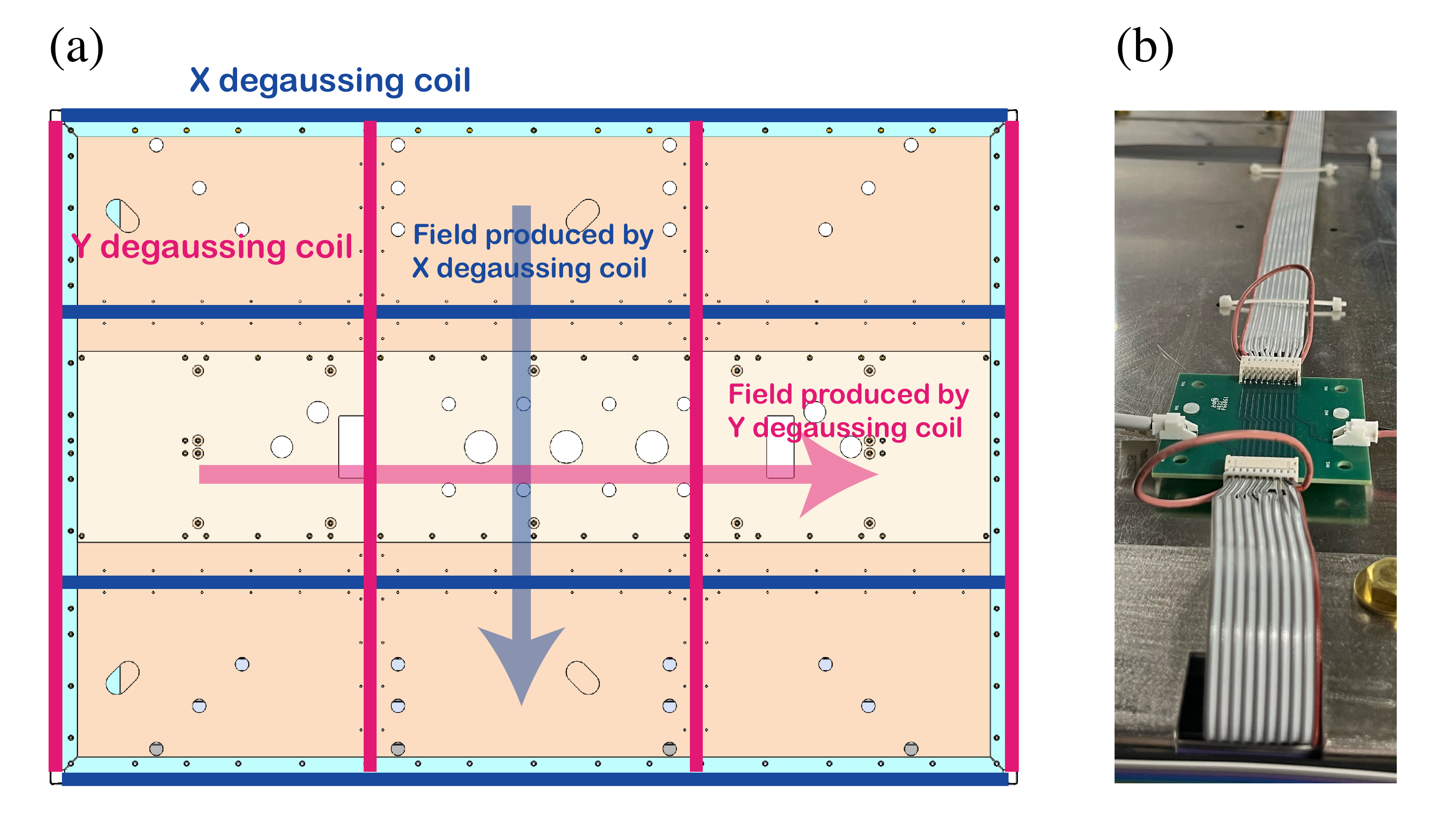}
  \caption{ The 8 coils that degauss one face of a shielding layer. The edge coils also  degauss the shield face ``around the corner."   The actual degaussing coil consists of a 10-turn ribbon cable, 2.54 cm wide, that is wrapped closely around the mu-metal sheet, and a custom PCB that makes a series connection between individual ribbon cable conductors.  }
  \label{fig:OneFaceDegaussing}
\end{figure}

For ACME III, the 18 faces of the 3 flat shielding layers each have 8 degaussing coils wound about them as shown in Fig.\ \ref{fig:OneFaceDegaussing}a. The 4 degaussing coils at the edge of each plate also help degauss the face around the corner. Each of the 108 degaussing coils is made using a 10-conductor ribbon cable (AWG24), JST S20B-PUDSS-1 connectors, and 108 custom printed circuit boards (Fig.\ \ref{fig:OneFaceDegaussing}b) to make a 10-turn coil that can be robustly assembled and disassembled as necessary.    

\begin{figure}[htbp!]
  \includegraphics[width=\linewidth]{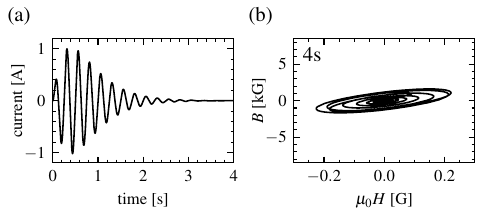}
     \includegraphics[width=\linewidth]{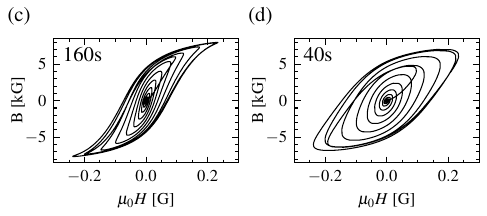}
    \caption{(a) A 4 second degaussing current pulse (most commonly used)  and (b) the resulting magnetic field within the mu-metal measured in-situ . (c) Expanding the pulse time scale to 160 s (d) and 40 s produce B vs.\ H curves that show more clearly the onset of saturation within the shields.}
  \label{fig:DegaussingWaveform}
\end{figure}

The degaussing current pulse used for all the degaussing coils has the form
\begin{equation}
I(t)=I_0\frac{t}{T}\exp\left(-\frac{t-T}T\right)\sin\left(2\pi f t\right). 
\label{eq:DegaussingWaveform}
\end{equation}  A 16-bit DAC (NI 9264) produces voltages at time steps of 0.05 ms.  After a $300$~Hz low pass filter, the voltage pulse is converted to a current (Kepco BOP 100-1M). Fig.\ \ref{fig:DegaussingWaveform}a shows the 4 second pulse we used most often, deduced from the voltage drop measured across a $1.0044\,\Omega$ shunt resistor. This shape is realized using $f = 4$ Hz and $T = 0.4$ s for the first 3.95 s, after which no current is sent to the degaussing coils. The peak current, $I_0 = 1$A, would produce a magnetic field of about $B \approx 0.25$~G if no mu-metal is present.  With the mu-metal present, this is a peak $H = B/\mu_0 = 20$ A/m (i.e. $\mu_0 H = 0.25$ G) that is much larger than the saturation value $H= 10$ A/m \cite{Arpaia2021} (i.e. $\mu_0 H = 0.125$ G).  

The magnetic field induced within the mu-metal shields is measured during degaussing by integrating the voltage induced across 10-turn pickup coils wound between holes in the shields spaced by 30 cm.  Fig.\ \ref{fig:DegaussingWaveform}b shows this measured field within one mu-metal layer, as a function of the applied field, as it is being degaussed with one of the 4 s degaussing pulses.  The area within the contours reduces as $H$ is slowly reduced.  Even though the 4 s pulse sufficed to realize a field in the precession volume of less than $10\,\uG$, saturation was not achieved.  Figs.\ \ref{fig:DegaussingWaveform}b-c shows the approach to saturation, close to the $8$ kG specification for the mu-metal \cite{CoNeticAA}.  The same peak current was used in these two cases, but the digitally generated pulses were expanded in time from 4 s to 160 s and 40 s (by scaling $T$ and $f$ in proportion) without changing the relative pulse shape.   

If offset currents in the degaussing coils are not carefully zeroed, a detectable remanent field will be present within the shields. The offset current from the bipolar current amplifier is constantly monitored by a 0.47 $\Omega$ precision shunt resistor and read out by a 24-bit NI-DAQ device. Prior to each degaussing pulse, the offset current was measured and reduced below $50\,\mu A$. The residual field can vary by as much as $2\, \uG$ from one identical degaussing cycle to another.  The residual field also varies by up to $5\,\uG$ depending upon which set of inner shields is degaussed last. These variations suggest that further optimization of the degaussing may be possible.

A complete degaussing process uses nine 4 s current pulses and takes 36 s. 
The first is sent to 12 degaussing coils wired in series to produce a field in one shield layer that circulates about the $x$ axis. The second and third pulses make magnetic fields that circulate about the $y$ and $z$ axes. The outer shield layer is degaussed first, then the middle and inner layers.  Our study indicates that among the outer, middle, and inner layers, the degaussing of the inner layer is the most important. In scenarios where the ambient magnetic field remains relatively stable, degaussing solely the inner layer yields results that satisfactorily fulfill the ACME III field requirements. Furthermore, in cases where only the  applied magnetic field within the shield's internal volume undergoes alterations, the degaussing of only the inner layer is adequate.


\section{Performance of the Coil-Plus-Shield System}
\label{sec:Performance}

\subsection{Magnetometers}
\label{sec:Magnetometers}

Three types of 3-axis magnetometers are used to measure the ambient magnetic fields, the fields produced by the coils, and the remanent field within the precession volume.  Table \ref{table:Magnetometers} displays their very different sensitivities and ranges according to the datasheets of these magnetometers. The datasheet sensitivities were verified by measuring the magnetic field noise. The magnetometer noise is the square root of the power spectral density (PSD) computed from the time dependent magnetic field signal. For all magnetometers, the magnetic field noise is slightly larger than specified in the datasheet, but  less than the $5 \, \uG$ that is needed.  However, offsets of the fluxgate and magnetoresistance sensors are as large as a few mG and are not very stable, which makes it difficult to measure the absolute field at the $\mu$G level. The  magnetism of the ethernet connectors used for the magnetoresistance sensor also complicates its use. However, both the fluxgate and magnetoresistance magnetometers can operate in the earth's field, but the Rb magnetometers cannot.

\newcommand\topspace{\rule{0pt}{2.6ex}}       
\newcommand\botspace{\rule[-1.2ex]{0pt}{0pt}} 

\begin{table}[h!]
    \centering
    \begin{tabular}{|l|rrr|}
    \hline
       {\bf Magnetometer}                      & ~~~~~~{\bf Sensitivity}    & ~~~~~{\bf Range}     & ~~~~~{\bf Specification } \\
         ~          & ~ $\uG/\sqrt{Hz}$       & ~              & ~~~~~{\bf Accuracy}\\
        \hline
        Fluxgate               & $0.05$               & $\pm 1$ G       & $0.5 \%$ \topspace \\
        Magnetoresistance~~~~~ & $3$                  & $\pm 1$ G     & ---~~ \\
        Rb                     & $0.00015$            & $\pm 500~\uG$       & ---~~  \\
        \hline
    \end{tabular}
    \caption{Specifications of the magnetometers. The fluxgate magnetometer is a Bartington Mag-13MSL100 or Mag-03MCTPL.  The magnetoresistance magnetometer is a Twinleaf VMR. The laser-pumped Rb magnetometer is a Quspin QZFM-Gen3 operated in zero-field mode. The accuracy of two of these is not specified.}
    \label{table:Magnetometers}
\end{table}

Magnetoresistance magnetometers are used to monitor the ambient magnetic field in the laboratory, illustrated in Fig. \ref{fig:AmbientB}. They are also used to measure the field produced by coils when they are outside of the magnetic shields.  An array of such sensors, located between the shield layers, are intended to watch for long-term changes in the outer shielding layers.  Laser-pumped Rb magnetometers measure the magnetic field in the precession volume when the magnetic shields are in place. 

The significant offsets of all the sensors are measured by rotating or reversing the direction of the sensors in a stable field.  The stable field environment was provided within the interior volume of  16.1 cm $\times$ 17.3 cm $\times$ 35.6 cm three magnetic layer shield, or within the much larger volume within the  3 layers of mu-metal shielding described in Sec.\ \ref{sec:MagneticShields}.   The largest offset correction for a magnetoresistance magnetometer was $\pm $ 3.5 mG, and these offsets typically varied by about  0.1 mG over a month.  The largest offset correction for a Rb magnetometers was 47 $\mu$G, and these offsets are observed to change by less than  $\pm $ 3$\mu$G over about 2 months. The fluxgate magnetometer output offsets ranged from $\pm$ 80-300 $\mu$G. 

Only the fluxgate magnetometers came with a specified scale accuracy (after offset correction). We built a 41 cm long solenoid with a 4 cm inner diameter to test the calibration. Two fluxgate magnetometers agreed with what we expected for the solenoid, and with each other, with an uncertainty that is a factor of two better than the specified 0.5\% accuracy.  The fluxgate magnetometers could then be used to calibrate the magnetoresistance and the Rb magnetometers. Corrections required for the magnetoresistance and Rb magnetometers sensors were performed by multiplying their native reading for the component corresponding to $B_z$ by 
factors of around 1.2.

\subsection{Coil Field (No Shields)}
\label{sec:CoilWithoutShields}

The performance of the actively shielded coil was evaluated first before it was enclosed within a magnetic shield. The magnetoresistance magnetometers were used because they are small, can operate close together without crosstalk, and they operate in fields up to 1 G field. To minimize the effect of time variations in the laboratory field, the coil was energized with a current $I = I_0 \cos({\omega t})$ that oscillated at 1 Hz and was measured using the voltage drop of a 1.004 $\Omega$ shunt resistor.  The oscillation of the magnetic field $B_z = B_{z0} \cos(\omega t)$ that was in phase with the drive current was measured throughout the precession volume, so magnetometer offsets that were stable over a measurement were removed.        

Measured fields along the $x$ axis are the points added to Fig.\ \ref{fig:CoilContributions}.  At the center, the measured and calculated value are 
\begin{align}
 B_z/I &=  257 \pm 2~\uG/\text{mA}~~~~~\text{\it measured}\\ 
  B_z/I &= 257 \pm 1 \,\uG/\text{mA}~~~~~\text{\it calculated}.
\end{align} 
The measurement uncertainty is primarily due to the scaling error of the magnetometer, the statistical fluctuations of magnetic field readings, and the uncertainty of the supplied current monitored across the shunt resistor.
The calculation uncertainty is due to the finite-element approximation of the differential equation, and the discrepancies of simulated geometries and actual apparatus because of the tolerance during the fabrication process.  The measured orthogonal field components are $|B_x/I|<1.7\,\mu\text{G/mA}$ and $|B_y/I|<3.0\,\mu\text{G/mA}$. 

The maximum observed spatial variation  of $B_z$ in the precession volume is $0.44 \,\mu\text{G/mA}$, or $0.18\%$ from the nominal value at the center. This compares favorably to the calculated homogeneity of $0.33\%$. For the nominal field of $100 \, \uG$, this is only $0.18 \,\uG$.

\subsection{Residual Magnetism of the Shields}
\label{sec:ResidualField}

The residual magnetic field within the spin-precession volume directly contributes to the non-reversing magnetic field during eEDM measurements. After a complete degaussing of the magnetic shields, the residual fields were measured using a Rb magnetometer  inserted into the shield through some of the many access holes in the shields.  A sliding rail system positioned the magnetometer with a position accuracy of approximately $\pm 1\,\text{mm}$. At each position, the magnetometer was rotated about the $x$ axis to determine and eliminate offsets in $B_y$ and $B_z$.  Magnetometer offsets in $B_x$ were determined and eliminated by reversing the probe direction in the sensor's $z$ axis.
Based upon the observed stability of the offsets, we can also measure the offsets within a small separate calibration volume at a different location, shielded by 3 mu-metal layers. These offsets can then be applied from a database as the measurements are being made.

\begin{figure}[htbp]
  \includegraphics[width=\linewidth]{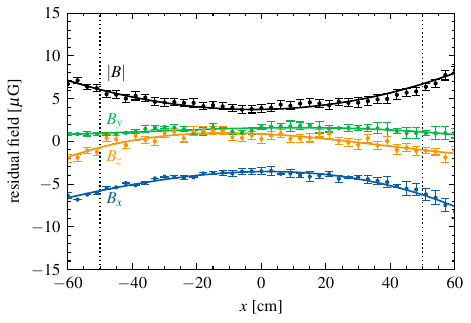}
   \includegraphics[width=\linewidth]  {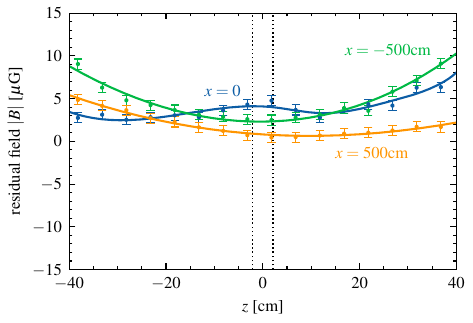}
   \caption{ (a) Components and magnitude of the residual field from the mu-metal shields along the central x axis and (b) transverse at 3 locations in x. Smooth curves are added to help guide the eye.}
  \label{fig:BResidual}
\end{figure}

The residual field along the central $x$ axis of the precession volume, 
measured with a Rb magnetometer, is shown in Fig.\ \ref{fig:BResidual}a.  The residual fields along the $y$ and $z$ axis are shown in Fig.\ \ref{fig:BResidual}b.  The residual fields are at or below $10 \, \uG$, and the gradients below $1\, \uG/\text{cm}$, throughout the precession volume. This is slightly better than the goal discussed in Sec. \ref{sec:ACMEIII}.  This low residual field within the shields, and within the ACME III vacuum chamber, was measured over the course of about a year.  During this time, the shields layers were removed and reassembled.

After these field measurements, additional ACME III apparatus was inserted into the vacuum chamber so eEDM measurements could be started.  After, the residual field that was measured was about a factor of three larger for reasons that are not clear. All materials inserted were carefully screened for their magnetism, and the array of magnetometers could not identify any localized source of the residual field. It may be that the magnetic shielding became a bit more magnetized the last time it was removed and reassembled.

The residual field increase is not a problem for ACME III because the increase in the residual field was simply canceled using the internal coils that were designed for this purpose. With this shimming, the residual magnetic field and gradient throughout the whole interaction volume were below the goal for ACME III.

\subsection{Shielding of Ambient Magnetic Fields}

\subsubsection{Normal Ambient Field}

After the shields are degaussed, the measured magnetic field in the precession volume is about $5 \,\uG$. This is a reduction by a factor of $10^5$ below a 0.5 G ambient magnetic field.  Normally, the ambient field varies by a small fraction of its average value, as illustrated in Fig.\ \ref{fig:AmbientB}b. We do not detect such fluctuations within the shielded precession volume.  

\subsubsection{Fringe Field from a Solenoid}

The fringe field of a high field solenoid at the shields changes between about $-260$ mG and $260$ mG (at the field monitor 2 m about the center of the precession volume) as the solenoid is ramped between $-5.5$ and $5.5$ T. This large change, comparable in size to the usual ambient field at the shields, provides a chance to test the shields.  

\begin{figure}[htbp]
 \includegraphics[width=\linewidth]{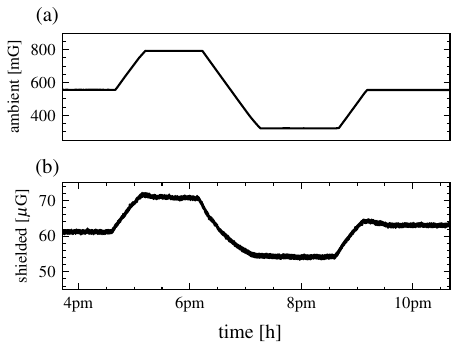}
 \includegraphics[width=\linewidth]{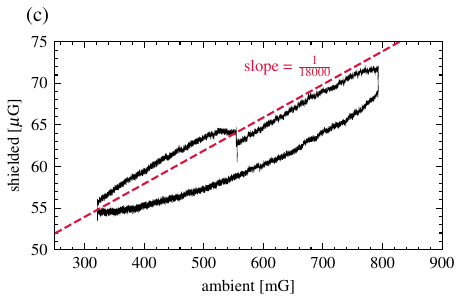}
  \caption{(a) $|\mathbf{B}|$ at a field monitor located 2 m above the center of the precession volume.  (b)  $|\mathbf{B}|$ at the center of the precession volume.  (c) Ratios of the measurements in (b) and (a). The shields were degaussed before this measurement but not during it.}
  \label{fig:ShieldSolenoid}
\end{figure}

The first test with the nearby external solenoid is to degauss the magnetic shields in the normal ambient field in the laboratory, then linearly ramp the nearby solenoid between its minimum and maximum field values.  Fig.\ \ref{fig:ShieldSolenoid}a shows the measured magnetic field components and magnitude at the ambient field monitor (2 m above the center of the shields).  Fig.\ \ref{fig:ShieldSolenoid}b shows the measured magnetic field components and magnitude at the center of the shielded precession volume.  To give some indication of the shielding that the mu-metal shields provide, Fig.\ \ref{fig:ShieldSolenoid}c relates the  measured shielded field to the field measured by the ambient field monitor.  For comparison, the slope of the dashed red line is a shielding of the unique solenoid fringe field by $1.8 \times 10^4$. Time delays in the response slightly reduce the correlation. 

The second test with the nearby external solenoid is to ramp it to full field, degauss the shields with a 4 s pulse used in Sec.~\ref{sec:Degaussing}, and then remeasure the residual field within the precession volume.  The field in the shielded precession volume does not change from what is shown in Fig.\ \ref{fig:BResidual}.  This indicates that the shielding factor of the ``new'' ambient field is $\gtrsim 2\times10^5$. Hence, we can adjust to a changing ambient field, even with the extremely large change from the nearby solenoid, as long as we degauss the shields once the ambient field is at a new stable value.  The strikingly larger shielding factor for truly static/ambient fields (that do not change after degaussing) than for quasi-static fields (that change, even slowly, after degaussing) has been similarly observed, and its mechanism explained, in Refs.~\cite{StaticVQuasi1,StaticVQuasi2}.

\subsubsection{Time Varying External Fields}

The shielding of an external field depends upon the fluctuation frequencies of the magnetic field applied at the outer boundary of the magnetic shields.  The permeability is frequency dependent, and eddy currents in the mu-metal, the aluminum supports, and the vacuum chamber can play a role.  In addition, the direction of the time-varying field and its spatial distribution at the shields, are important.

Nonetheless, to get some idea how external time varying fields penetrate to the precession volume, we constructed 3 pairs of external coils just outside the outer magnetic shield.  Each pair produced a magnetic field in either $x$, $y$ or $z$ directions. The field inside the shields was measured with a Rb magnetometer at the center of the precession volume. The currents in the coils were varied sinusoidally at various frequencies and the in-phase component of the detected field was extracted.   Fig.\ref{fig:ExternalCoils} gives an example of how the shielding factor changes with frequency.   

\newcommand{\Bext}{\mathbf{B}_\mathrm{ext}}

\begin{figure}[htbp]
  \includegraphics[width=\linewidth]{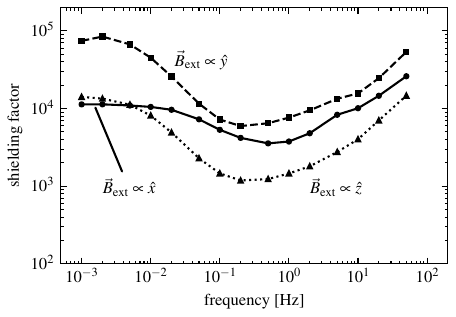}
  \caption{Shielding factor for external fields $\Bext$ applied outside the mu-metal layers in the  $\xhat$, $\yhat$ and $\zhat$ directions.  }
  \label{fig:ExternalCoils}
\end{figure}

The observed shielding decrease out to $0.1$ Hz is primarily attributable to the reduced magnetic permeability of Co-Netic AA. The increase in shielding above 1 Hz likely comes from eddy current shielding.  For all directions and all frequencies the shielding reduction is greater than $10^3$.

\subsection{Coil-Plus-Shields System}
\label{sec:CoilWithShields}

\subsubsection{Field Calibration and Uniformity}

Enclosing a coil within a magnetic shield will change the field produced by the coil, especially if the shield is close to the coil, because the shield imposes a very different boundary condition than free space.     Using a Rb magnetometer at the center of the precession volume, we measure
\begin{equation}
B_z/I = 258 \pm 1 ~\uG/\text{mA}.   
\end{equation}
The uncertainty comes from the scaling error of the magnetometer, the statistical fluctuations of magnetic field readings, and the uncertainty of the supplied current monitored across the shunt resistor.
This agrees with the ratio measured with no shields present within the measurement uncertainty.  This agreement is a manifestation of how well the coil and the shields are decoupled because of the actively-shielded coil design. We shall see an example of a 30\% change in field in App.\ \ref{sec:AuxCoils} for a coil that is not actively shielded.   

\begin{figure}[htbp!]
    \centering   
    \includegraphics[width=3.5in]{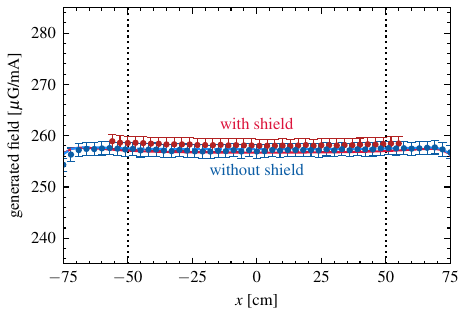}
    \centering   
    \caption{The measured and calculated fields with and without surrounding magnetic shields agree within their uncertainties.} 
    \label{fig:CoilWithAndWithoutShield}
\end{figure}

The spatial homogeneity of the field produced by the coil is also measured to be the same with and without the shields present. Fig.\ \ref{fig:CoilWithAndWithoutShield} compares the measured field and calculated fields, with and without a nearby shield surrounding the actively-shielded coil.  There is good agreement within the uncertainties.  

\subsubsection{Testing for Induced Magnetization}

\begin{figure}
    \centering
     \includegraphics[width=\columnwidth]{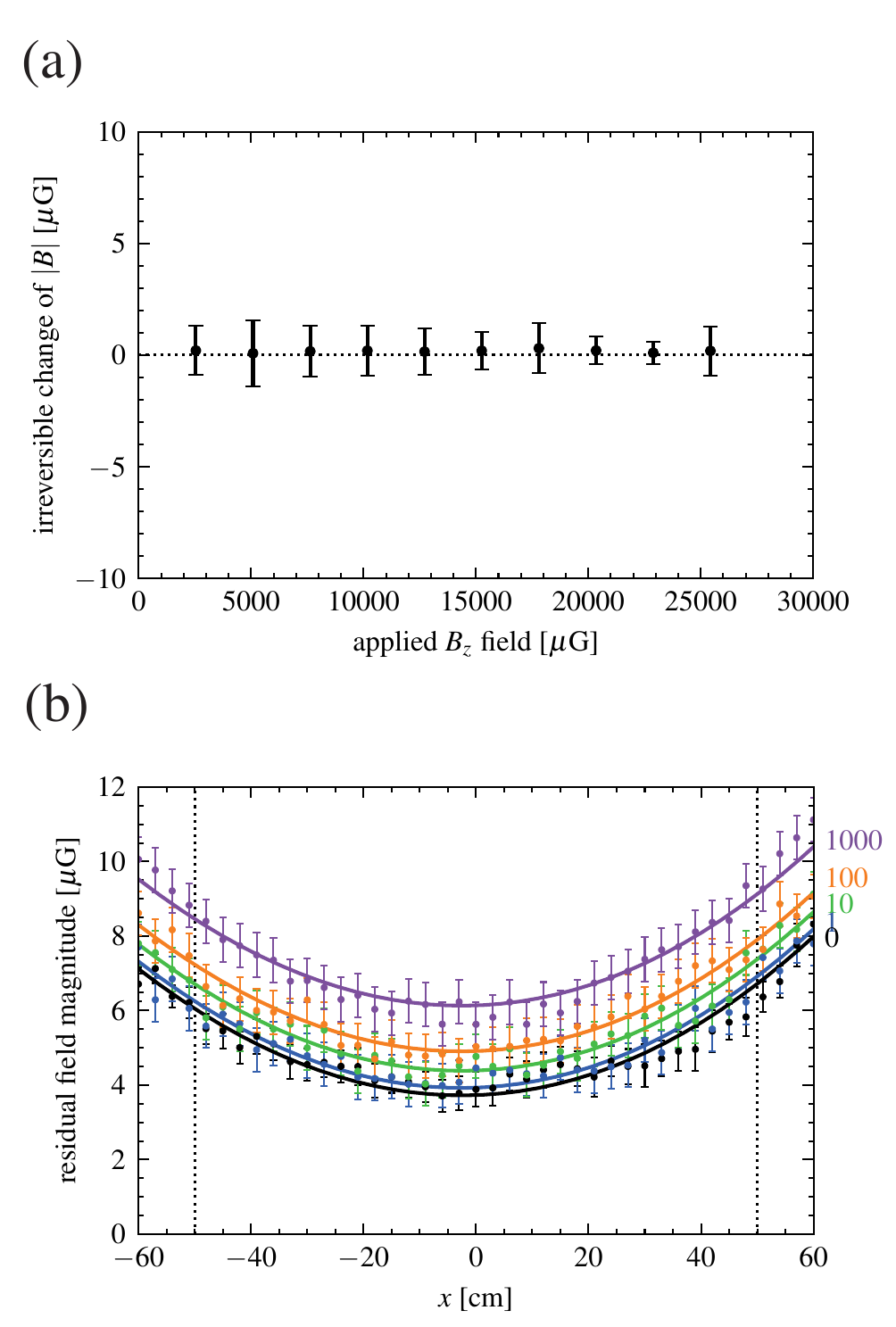}
    \caption{(a) The measured change of residual field at the center of precession region after applying a 25 mG field. (b) The residual field measured on the central x axis, with various number of cycles of applying reversing Bz field of 2 mG. }
    \label{fig:MagnetizationBuildup}
\end{figure}

As a first test, the residual magnetism within the shielded precession volume is measured before and after magnetic fields were produced by the coil.  The magnetization is a concern because the nearest shield is only about 10 cm away from windings in the coil, and because such magnetism had been important for ACME II. 
The shields were degaussed before the test but not during. Fig.\  \ref{fig:MagnetizationBuildup}a shows the difference of the residual field measured at the center of the precession volume before and after the coil produces a magnetic field.  The strength of the applied field increases from left to right up to about 25 mG.  No increase in shield magnetism is detected. The actively-shielded coil was designed to minimize the fringing field at the nearest shield to minimize possible shield magnetization.  Applying much larger than expected fields produces no detectable magnetization, even by a field that is 250 times bigger than the nominal field to be used.      

A second test is for residual shield magnetism after the direction of the field produced by the current is repeatedly reversed.  The shields were degaussed before but not during the complete test.  The lower black points in Fig.\ \ref{fig:MagnetizationBuildup}b shows the field along the center $x$ axis of the precession volume after degaussing.  A field of 2 mG (20 times larger than the nominal $100 \,\uG$ that is the nominal field for ACME III) is then applied and reversed in direction every 30 seconds (the fastest expected reversal time for ACME III). The other curves in  Fig.\ \ref{fig:MagnetizationBuildup}b are taken after 10, 100 and 1000 cycles of switching the field direction and then restoring it.  The result suggests that magnetization field does build up, but it is about 5 $\uG$ for 1000 reversing cycles of the field, which is about 17 hours. This encouraging result suggests that it may even be possible to make measurements over many hours of field switching without the need to spend valuable measurement time degaussing.

A third test is to degauss the shields, and then measure the residual field from the shields before and after a 25 mG field was applied for 24 hours. No increase in the shield magnetism was observed at the $1 \, \uG$ level.  

\subsection{In-situ Co-magnetometry}
\label{sec:comag}

The magnetic field produced and measured with the Rb magnetometers can be checked with entirely different sources of measurement uncertainty by using a beam of ThO molecules.  One of the three components of the magnetic field, $B_z$, can be measured, averaged over the 1 m x 4.2 cm x 4.2 cm spin precession region.   It is also possible to use ThO to measure changes in strength of two of the possible field gradients,  $\partial B_z/\partial z$ and $\partial B_z/\partial y$, also averaged over the spin precession volume.   A comparison provides a valuable consistency check on the Rb magnetometer measurements, and the ThO co-magnetometers, and upon the uncertainties in these measurements.  

Spin precession measurements on the metastable Q $^3\Delta_2$ state of thorium monoxide are used.  In contrast to the H $^3\Delta_1$ state used for the EDM measurement, the Q state has a large magnetic dipole moment, $g_Q\approx2$, and has no sensitivity to the electron EDM \cite{Wu2020}. A new measurement of the ThO Q state g-factor is shown in \ref{sec:Q state dipole moment}. The ThO beam is produced by pulsed laser ablation of a solid ThO$_2$ target in a cryogenic buffer gas beam source, primarily in the ground state X. After rotational cooling to the lowest rotational state, the molecules are transferred to a superposition $(\ket{Q, J=2, M=2, \Omega =-2}+\ket{Q, J=2, M=-2, \Omega =2})/\sqrt{2}$ via STIRAP through the intermediate C state. The molecules are then focused by the electrostatic lens before flying through a region with no applied electric field or magnetic field, allowing the state to freely mix within the Q state manifold. The resultant state at the beginning of the interaction region is an incoherent mixture of all sub-levels of the Q, J=2 manifold.

\begin{figure}
    \centering
     \includegraphics[width=\columnwidth]{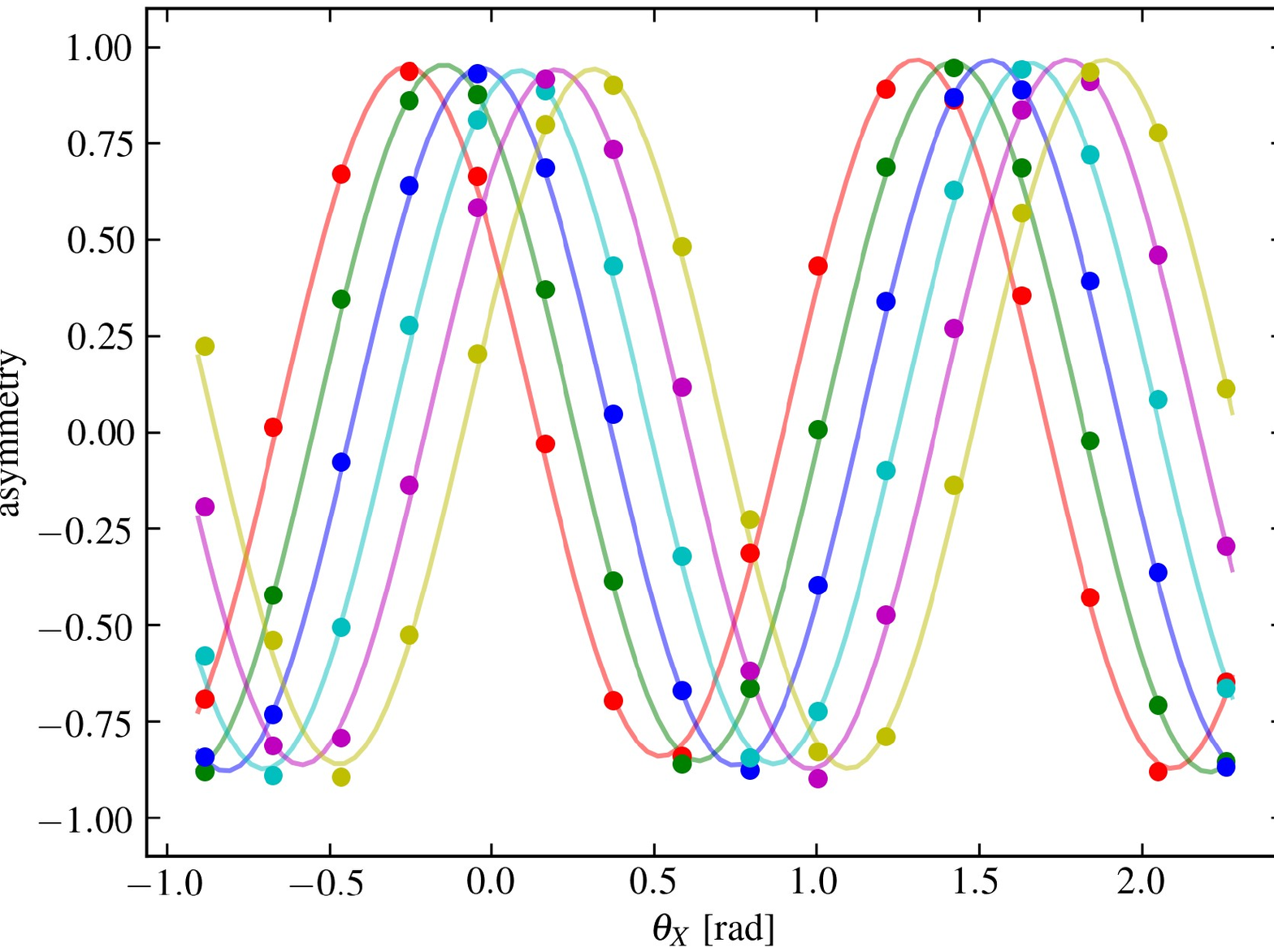}
    \caption{Measured asymmetry fringes with 8 $\mu$G increments. Uncertainties on the asymmetry points are smaller than the displayed marker.}
    \label{fig:asymmetry fringes}
\end{figure}

Inside the interaction region, the molecule is polarized in a 160 V/cm electric field applied by glass field plates coated with indium tin oxide. An x-polarized laser resonant with the $\ket{Q, J=2, M=\pm1, \Omega=\pm1}\rightarrow\ket{I, J=1, M=0, P=-1}$ illuminates the molecules near the beginning of the interaction region. The interaction of the laser with the molecule results in a bright state and a dark state defined by the laser polarization. The bright state is rapidly pumped away, leaving only molecules in the dark state $(\ket{Q, M=+1, \Omega=+1}+\ket{Q, M=-1, \Omega=-1}) / \sqrt{2}$. The molecules then freely precess during a 5 ms flight time through the interaction region. The M=+1 and M=-1 states experience opposite Zeeman shifts such that at the end of the interaction region the state evolves to $(e^{-i \phi}\ket{Q, M=+1}+e^{i \phi}\ket{Q, M=-1}) / \sqrt{2}$ where $\phi=\mu_Q \bar{B}_z \tau/\hbar$ and $\bar{B}_z$ is the average magnetic field experienced by the molecule in the interaction region. A readout laser is rapidly switched at 250 kHz between orthogonal polarizations with acousto-optic modulators, and the fluorescence is recorded on eight Silicon Photomultiplier modules. In each polarization bin, the asymmetry is calculated from the fluorescence signal from the two polarizations $A=(S_X-S_Y)/(S_X+S_Y)$. After recording the fluorescence, the angle of these laser polarizations relative to the lab x axis is rotated with an achromatic half waveplate over 180 degrees. The asymmetry is of the form
\begin{equation}
A=C\cos[2(\phi-\theta_X)]
\end{equation}
where $C=0.90\pm0.01$ is the contrast and $\theta_X$ is the angle of the X laser polarization with respect to the lab x axis. The asymmetry is fit to a cosine to extract the phase difference. In order to measure the magnetic field, the laser polarizations of the prep and probe lasers must be aligned. We take advantage of the molecular beam velocity dispersion to align these in-situ: the velocity of the ThO beam varies by about 10\%, with the fastest molecules arriving first. Slower molecules have a greater coherence time and thus a greater phase. The spread in the measured phase is therefore proportional to the absolute magnetic field. The laser polarization corresponding to maximal asymmetry when the spread in phase across the beam is minimized is the aligned state. By fitting a line to the spread in the molecule phase as the magnetic field is stepped in 4 $\mu$G increments around 0 $\mu$G, the polarizations are aligned to 9 mrad. This gives a systematic uncertainty in the magnetic field measured with co-magnetometry of 150 nG.

\begin{figure}
    \centering
     \includegraphics[width=\columnwidth]{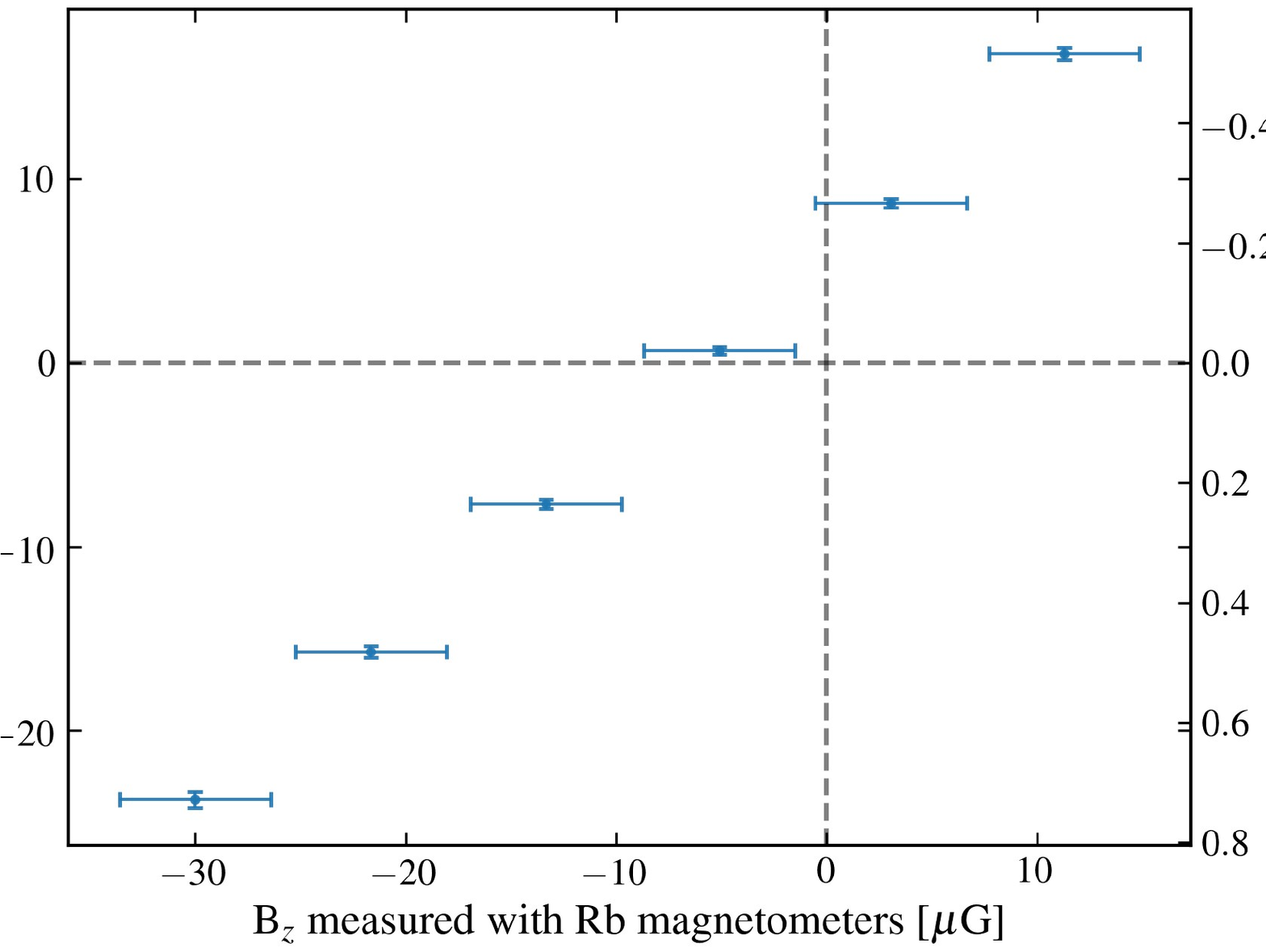}
    \caption{Comparison of the measured $\mathrm{B}_z$ field with co-magnetometry versus with the Rb magnetometers located in pockets}
    \label{fig:Bz_comag}
\end{figure}

As a cross-check of the co-magnetometer measurements, we compare to the field measured with eight Rb magnetometers inserted in the side pockets. We approximate the field with a second order Taylor series expansion about the center of the interaction region. The average field, and first and second order gradients are derived by performing a least-squares minimization fit of the fifteen magnetic field values measured by the Rb magnetometers. Maxwell's equations $\nabla\cdot B = 0$ and $\nabla \times B = 0$ are enforced in the fit.

To compare the field measured with the magnetometers in pockets versus the field measured with co-magnetometry, we apply several offset $B_z$ fields in 8 $\mu$G increments, shown in Fig.\ \ref{fig:Bz_comag}. The uncertainty in the $B_z$ is computed in quadrature from three sources. First, an estimate of the uncertainty from the position of the magnetometers is computed by simulating the variation in measured $B_z$ if the positioning of the magnetometers changed by 1 mm in the x or y directions, and 5 mm in the z direction. This gives a small uncertainty of 120 nG. Next, the same simulation is done for the uncertainty in the magnetometer offset calibrations which we estimate to be 3 $\mu$G, resulting in a 1.3 $\mu$G uncertainty. Finally, we take the mean of the absolute value of residuals, 2.6 $\mu$G, of the least-squares fit as an additional error which reflects our uncertainty about any small higher order gradients not included in the fit. The error bar for the co-magnetometry measurement is derived from the uncertainty in the polarization offset between the prep and readout (150 nG), together in quadrature with uncertainty about the precession time due to variation in the molecular beam forward velocity. The approximate precession time variation is 1.6\% and scales directly to give a 1.6\% uncertainty on the $B_z$ measured with co-magnetometry.

\begin{figure}
    \centering
     \includegraphics[width=\columnwidth]{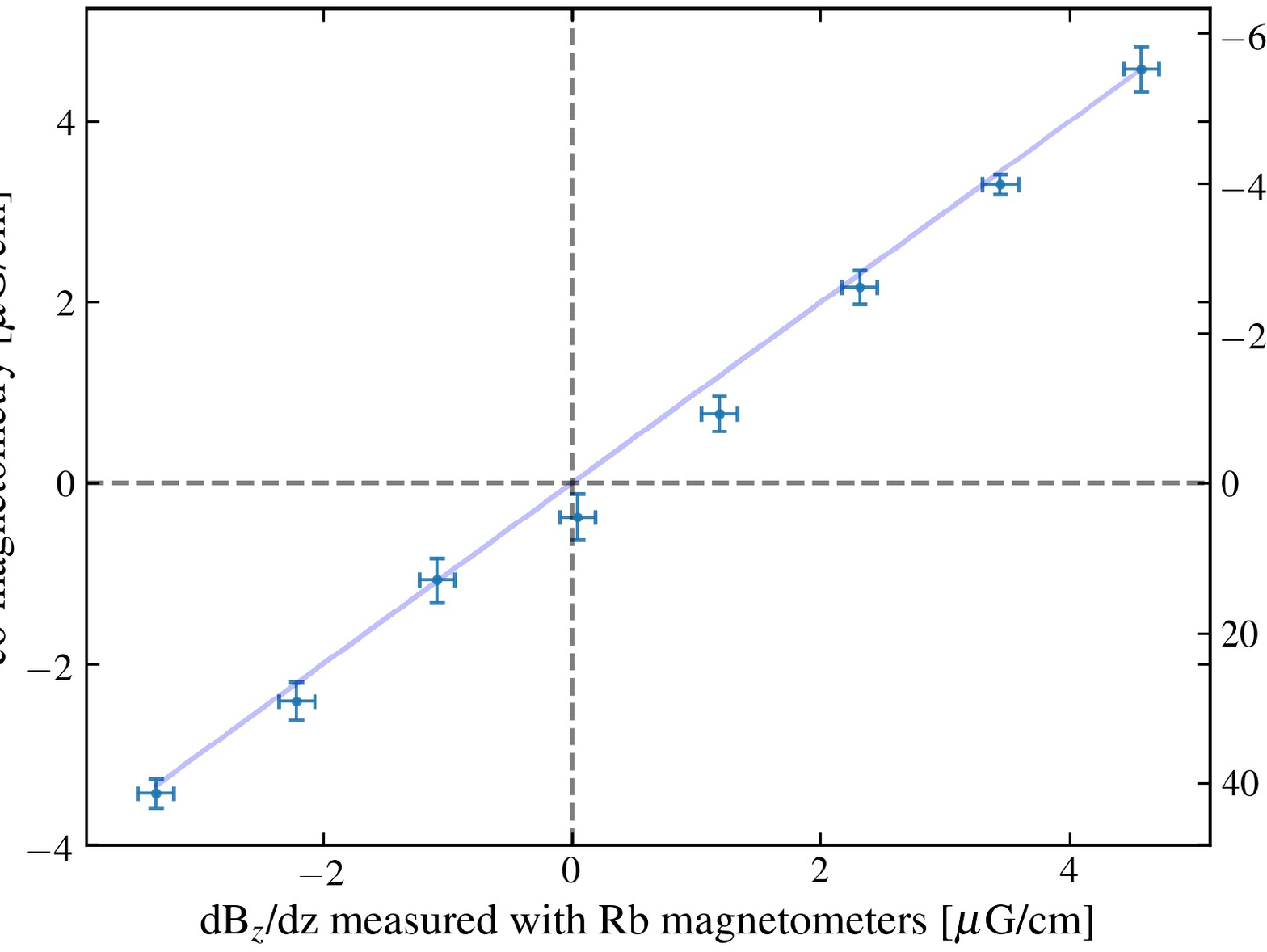}
    \caption{Comparison of the measured $\partial\mathrm{B}_z/\partial z$ field with co-magnetometry versus with the Rb magnetometers located in pockets. The line is a guide for the eye that shows the line of perfect agreement.}
    \label{fig:dBzdz_comag}
\end{figure}

Two first order gradients, $\partial B_z/\partial z$ and $\partial B_z/\partial y$ can also be measured with co-magnetometry. Consider that in the presence of a positive $\partial B_z/\partial z$ field, molecules with a positive average z position will accumulate a greater Zeeman phase than those with a negative average z position. The phase and magnetic field that we measure during $B_z$ co-magnetometry is the spatial average of these individual molecule phases. We measure gradients by varying the spatial distribution of molecules we detect and computing the difference with the full molecule distribution. Concretely, the correspondence between the measured phase difference and the magnetic field gradient is

\begin{equation}
\frac{\partial B_z}{\partial z}=\frac{\hbar\Delta\phi}{\mu_Q \tau\thinspace\Delta z_{CM}}
\end{equation}
where $\Delta z_{CM}$ is the shift in the center-of-mass of the detected molecular cloud. To create this center of mass shift in the z direction, we illuminate the molecules after the electrostatic lens and before the interaction region with a laser along y, resonant with the Q$\rightarrow$I transition, to optically pump molecules into the dark X state. A razor blade mounted to a motorized translation stage can be extended to block the laser beam at various points. To measure the gradient along z, we measure the molecule precession phase twice; once with the full molecular cloud, and once with the +z half of the molecular cloud pumped out of Q. For measuring the gradient along y, the readout laser beam is similarly clipped with a razor blade mounted to a motorized translation stage along y. The laser is alternately fully unblocked or clipped to block the -y half of the molecular beam and the difference in phase is measured. For both gradients, since the center of mass shift depends on the position-velocity correlation of the molecule cloud, the conversion of $\Delta \phi$ to $\partial B_{z}/\partial z$ or $\partial B_{z}/\partial y$ must be calibrated with a known magnetic gradient change which we apply with the $\partial B_{z}/\partial z$ and $\partial B_{y}/\partial z$ coils, respectively.

\begin{figure}
    \centering
     \includegraphics[width=\columnwidth]{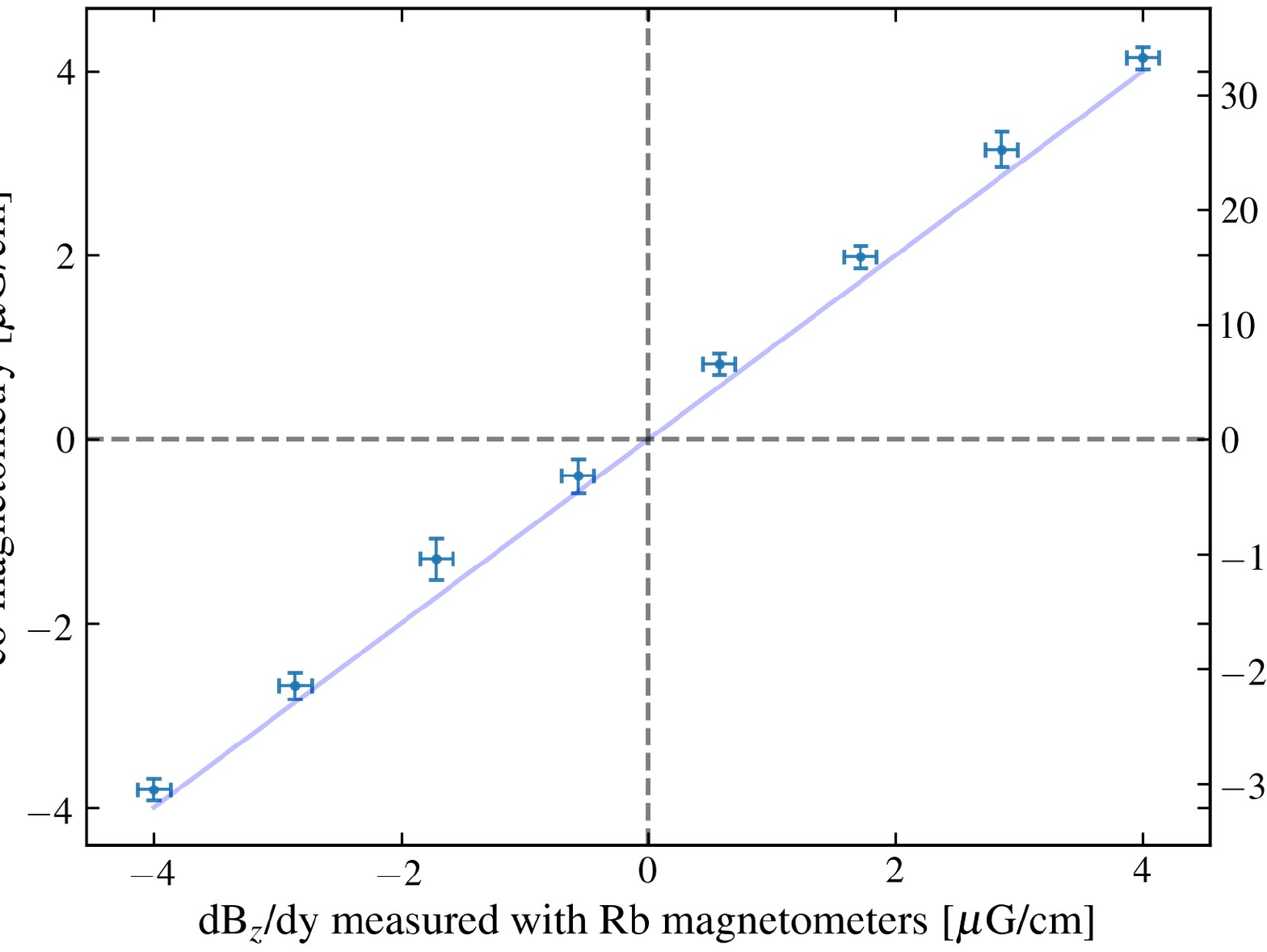}
    \caption{Comparison of the measured $\partial\mathrm{B}_z/\partial y$ field with co-magnetometry versus with the Rb magnetometers located in pockets. The line is a guide for the eye that shows the line of perfect agreement.}
    \label{fig:dBzdy_comag}
\end{figure}

The gradients measured via co-magnetometry are compared to those from the Rb magnetometers in Fig.\ \ref{fig:dBzdz_comag} and Fig.\ \ref{fig:dBzdy_comag}. The uncertainty in the measured values from the pocket magnetometers is derived from the same three sources as for the $B_z$ measurement, except that the contribution computed from the residuals is divided by the characteristic distance between magnetometers along y or z, respectively. The uncertainty in the co-magnetometry measurement is from the fluctuation in repeated measurements at the same applied gradient field. These two methods show good agreement to within 200 nG/cm. 

\section{Conclusion}

An actively-shielded coil, surrounded by three layers of
mu-metal shielding, provides a uniform magnetic field over a 1 meter long precession volume. The magnetic shielding keeps the ambient magnetic field and its fluctuations from affecting the measurement. The small fringing field, a consequence of the active shielding design, barely magnetizes a mu-metal shield only 10 cm away, without requiring frequent degaussing of the shields. The component of the field that does not reverse when the coil current is reversed is small enough to not add error to the projected statistical uncertainty of a ten times more sensitive eEDM measurement. Based upon the demonstrated performance, a new ACME III measurement with more than an order of magnitude improvement in eEDM sensitivity should be possible using the new coil-plus-shields system.

\section*{Acknowledgments}

This work was supported by the National Science Foundation, the Gordon and Betty Moore Foundation, and the Alfred P. Sloan Foundation. This work was performed as part of the ACME collaboration.


\appendix

\section{Auxiliary Coils}
\label{sec:AuxCoils} 

\begin{figure}[hbtb!]
\includegraphics[width=\the\columnwidth]{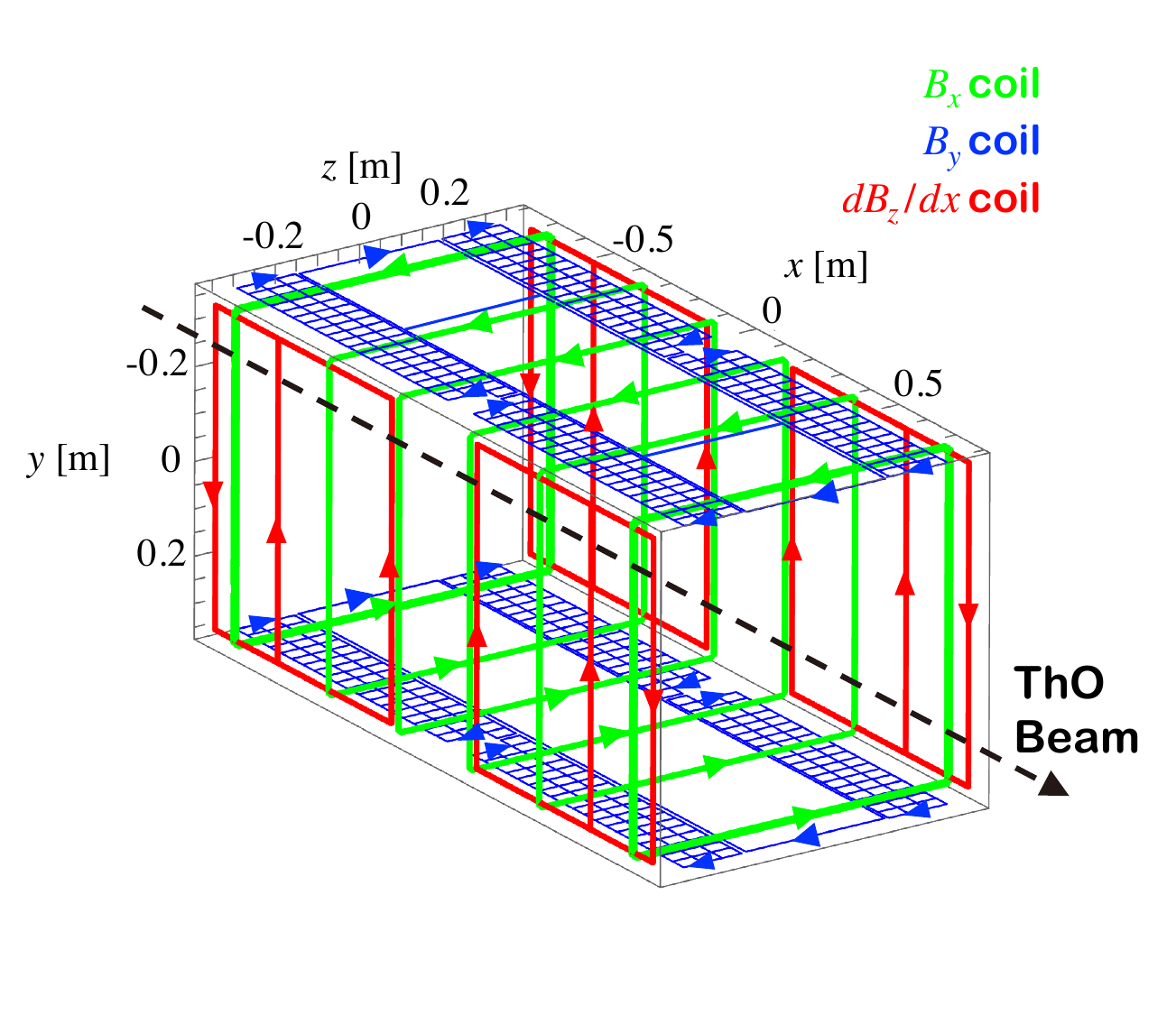}
\caption{Auxiliary coils are wound on the inner frame of the actively-shielded coil shown in Fig.~\ref{fig:CoilPicture}.}
\label{fig:AuxCoils}
\end{figure}

The actively-shielded coil produces a spatially uniform magnetic field in the z direction.   
Fig.\ \ref{fig:AuxCoils} shows additional coils, not actively shielded, that are added to the coil assembly to generate magnetic fields in the x and y directions.  They also produce 5 independent linear gradients.  (There are 9 possible linear gradients and 4 constraints from Gauss' Law and Ampere's law.)
These fields will be used to check for systematic errors.  If needed, they are also available to apply small fields and gradients to improve the spatial quality of the magnetic field in the precession volume.

\newcommand{\dB}[2]{\partial B_{#1}/\partial #2}

\subsection{``Green Coils''}

The green coils in Fig.\ \ref{fig:AuxCoils},  can produce either $B_x$ or the gradient $\partial B_x/\partial x$. There are 8 single turn loops, symmetrically located with respect to the center of the precession volume, made with 18 AWG wire pressed into 1/16 inch grooves. The two turns on each end cannot be distinguished in the figure because they are only separated by 1 cm. High density polyethylene (HDPE) clamps hold the wires securely in place.  The current in each of the 8 turns can be represented as a current vector, $I = \{I_1,\cdots,I_8\}$, where the indices represent the coils in order with respect to the $x$ direction along which the ThO molecules are traveling. 

To generate the most uniform $B_x=1\,\uG$ field, the optimized current vector is 
\begin{align}
I &= \{23.2,20.2,22.1,20.6,20.6,22.1,20.2,23.2\}\,\mu A.
\end{align}
These currents can be linearly scaled to produce other values of $B_x$. The nominal field is at the center of the spin precession region. Within the spin precession volume, the maximum field deviation from the nominal value at its center is about 0.96\%. 

\begin{figure}[htbp]
    \centering
   \includegraphics[width=\linewidth]{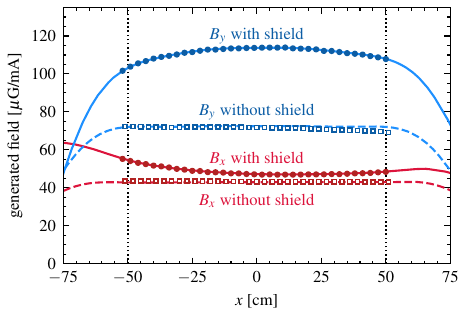}
\caption{The magnetic field produced by the green coil in Fig.\ \ref{fig:AuxCoils} on the central $x$ axis in the $x$ (red) and $y$ (blue) directions.  The measured points and calculated curves are consistent when the surrounding magnetic shields are not (dashed) and are (solid) present.} 
    \label{fig:AuxBx}%
\end{figure}

Fig.\ \ref{fig:AuxBx} shows the good agreement between measured and calculated values of $B_x$ along the central x axis of the precession volume.  The field values without shield were calculated using the Biot-Savart law, assuming free-space surroundings.   The field values when shields are present were calculated using COMSOL finite element analysis software using a physical model of the shielding material. For both the $B_x$ field and $B_y$ field (which will be discussed in detail in Sec. \ref{sec:AuxCoils} c), the presence of the enclosed magnetic shield significantly increases the generated field, consistent with the concept that high-permeability shield materials work as the return yoke for the coil. The pronounced 12\% increase caused by the shields is because this coil does not have the active shielding that greatly reduces the coupling between the coil and the shield. By contrast, the shift for the actively shielded coil shown in Fig. \ref{fig:CoilWithAndWithoutShield} is barely detectable. The largest deviation for $B_x$ in the precession volume, from the nominal field value at the its center, is about $17.2\%$ . The field distribution is affected by the surrounding shields.

The optimized current vector for producing $\dB{x}{x}  = 1\, \uG/\text{cm}$,
\begin{align}
I &= \{-1.89,-1.89,-0.849,-0.302,\\
&~~~~~~~~~0.302,0.849,1.89,1.89\}\,mA,
\end{align}
can be scaled linearly to produce other values of the gradient. Since $\grad \cdot \textbf{B} = 0$ and the coil is invariant under rotations by $\pi/2$ about the $x$-axis, this coil simultaneously produces the gradients
\begin{align}
    \dB{x}{x} = -2 \dB{y}{y} = -2 \dB{z}{z} = 1\, \uG/\text{cm}.
    \label{eq:dBxdx}
\end{align}
The measurement in Table \ref{tab:gradient_components} agrees within the measurement uncertainty. Over the entire spin precession volume, these gradients vary from the nominal value at the center by a maximum of 3.6 \%.  

\begin{figure*}[htbp]%
    \centering
   {{\includegraphics[width=6in]{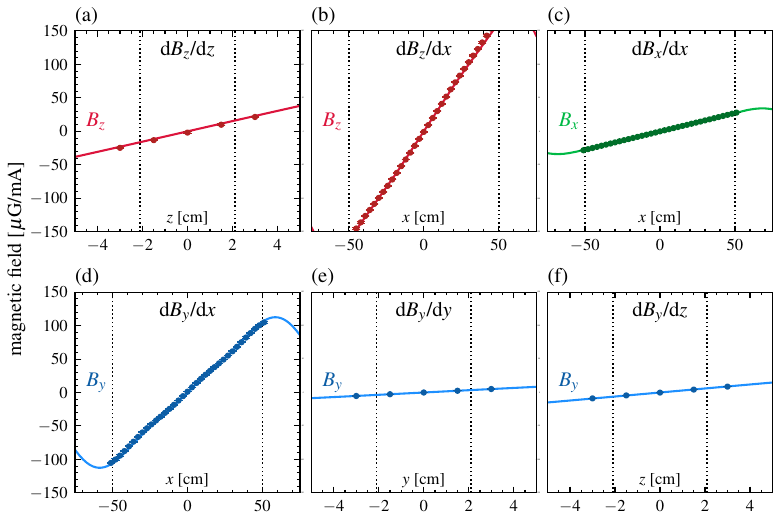} }}%
    \caption{The various gradients producible by the ACME III coil system. The extent of the spin precession region for each corresponding coordinate is shown as vertical dashed lines}%
    \label{fig:Gradients}
\end{figure*}

For both the uniform and gradient fields, the magnitudes of the currents sent through windings equidistant from the center are equal.  We thus use a 4 channel current supply (GPP-4323) with relays that can switch the direction of the current.  The good agreement between the calculated and measured gradients is shown in Fig.\ \ref{fig:Gradients}c.

\subsection{``Red Coils''}

The red coil in Fig.\ \ref{fig:AuxCoils} consists of four sets of rectangular coils parallel to the $xz$-plane, made using 24 AWG enameled wire pressed into 1/8 inch grooves.  Each set is comprised of an overlapped larger and smaller coil in the same plane. All loops are connected in series and a single current $I$ is sent through all of the windings, in the directions indicated by the arrows in the figure. Constrained by Ampere's law, this coil simultaneously produces two gradients, 
\begin{equation}
\frac{\partial B_z}{\partial x} = \frac{\partial B_x}{\partial z} = 3.2~\uG/\text{cm} \times I~(\text{mA}),
\end{equation}
where $I$ is the current from a single current supply (CS-580) sent through all the wires.  Within the precession volume, the calculated gradients vary from the nominal value at the center by a maximum of 20.1\%.  The good agreement between the calculated and measured gradients is shown in Fig.\ \ref{fig:Gradients}b.

\subsection{``Blue Coils''}

The blue coils are wound as a series of square cells, made with 24 AWG enameled wire in 1/8 inch grooves.  The winding details are in \cite{AngThesis2023}. The cells are wired as 14 separate circuits so that the current vector is $I = \{I_1, \cdots, I_{14}\}$.  The first 8 currents are the lattice coils. The order starts at the top left lattice coil in Fig .\ \ref{fig:AuxCoils} and moves counterclockwise then to the bottom set of lattice coils. Currents 9 and 10 are coils located at the center. Currents 11 through 14 are the four side coils. As a simplification, the cells are wound such that the current vector has the form $\vec{I} = I_0 \, \hat{i}$, where $\hat{i}$ has components of +1, -1, or 0 to designate one current direction, the other current direction, and no current flowing, respectively.  

Producing a field $B_y = 1 \, \uG$ requires 
\begin{align}
I = 12.3 \,  \mu \text{A} ~ \{1, 1, 1, 1, 1, 1, 1, 1, 1, 1, 0, 0, 0,0\}.
\end{align}
Different values of $B_y$ can, of course, be produced by linearly scaling all the currents.
Within the precession volume, the calculated field deviates from the nominal value at the center by about 0.71\%.  Fig.\ \ref{fig:AuxBx} shows the good agreement between measured and calculated values of $B_y$ along the central $x$ axis of the precession volume.  The lower values, for no surrounding magnetic shields, were calculated using the Biot-Savart law.   The higher values, including surrounding magnetic shields, were calculated via finite element analysis using COMSOL. When shielded, the maximum deviation of $B_y$ from the nominal value at the center of the precession volume is about $10.7\%$.  The large 50\% difference between shields present and not present is because this coil does not have the active shielding that greatly reduces the coupling between the coil and the shield.  (By contrast, we have seen that the shift for the actively-shielded coil that produces $B_z$ is barely detectable.)

Gradient fields can also be generated.  The two gradients, 
\begin{equation}
\dB{y}{x} = \dB{x}{y} =  1\,\uG/\text{cm},     
\end{equation} are simultaneously produced by the currents
\begin{align}
I = 0.476\, \text{mA} ~ \{-1, -1, 1, 1, -1, -1, 1, 1,0,0,-1,-1,1,1\}.
\end{align}
Within the precession volume, the maximum deviation from the calculated gradients vary from the nominal value at the center by about 31.6 \%.  The good agreement between the calculated and measured gradients is shown in Fig.\ \ref{fig:Gradients}d.

The two gradients,
\begin{equation}
\dB{y}{z} = \dB{z}{y} = 1\,\uG/\text{cm}
\end{equation}
are simultaneously produced using the currents
\begin{align}
I = 0.333 \,\text{mA} \{1, -1, 1, -1, 1, -1, 1, -1, 0,0,0,0,0,0\}.
\end{align}
Over the precession volume, the  calculated gradient differs from the nominal value at the center by as much as 4.8 \%.  The good agreement between the calculated and measured gradients is shown in Fig.\ \ref{fig:Gradients}f.

\begin{table}[htbp]
\centering
\begin{tabular}{|c|c|c|c|}
\hline \textbf{gradient coils} & d$B_x$/d$x$ & d$B_y$/d$y$ & d$B_z$/d$z$ \\
\hline \textbf{d$B_x$/d$x$ coil} & 0.55 & -0.31 & -0.26 \\
\hline \textbf{d$B_y$/d$y$ coil} & -0.41 & 1.81 & -1.40 \\
\hline \textbf{d$B_z$/d$z$ coil} & -0.45 & -7.35 & 7.78 \\
\hline
\end{tabular}
\caption{Measured gradients simultaneously generated by d$B_x$/d$x$, d$B_y$/d$y$ and d$B_z$/d$z$ coils. Units are in $\mu$G/(cm$\cdot$mA)}
\label{tab:gradient_components}
\end{table}

Producing $\dB{y}{y} = 1\,\uG/\text{cm}$ also produces
$\dB{x}{x}$ and $\dB{z}{z}$ such that 
\begin{align}
\dB{y}{y} = -\dB{x}{x} - \dB{z}{z}.
\label{eq:dBydy}
\end{align}
Given that the extent of the blue coils on $x$ coordinate is roughly $2.6$ times as it is on $z$ coordinate, we expect that the $\dB{z}{z}$ produced in this coil is roughly $2.6$ times $\dB{x}{x}$. The directly calculated values, confirmed with real measurements as shown in Table \ref{tab:gradient_components}, are:

\begin{align}
\dB{x}{x} &= - 0.23 \, \dB{y}{y}\\
\dB{z}{z} &= - 0.77 \, \dB{y}{y}.\\
\end{align}
The negative of their sum does give $\dB{y}{y}$ to satisfy Eq.\ \ref{eq:dBydy}. 

The required currents are 
\begin{align}
I = 0.556 \, \text{mA} ~\{-1,-1, -1, -1, 1, 1, 1, 1, 1, 1, 0, 0 ,0, 0\}.
\end{align}
Over the precession volume, the  calculated gradient varies from the nominal value at the center of this volume by as much as 22.1\%.  The good agreement between the calculated and measured gradients is shown in Fig.\ \ref{fig:Gradients}e.

Each of the field and gradient configurations uses the current from a single source (CS-580) sent through all the windings in series.  The current directions are selected by relays. A possible upgrade path is to send 14 different currents though the 14 windings to slightly increase the spatial quality of the field and gradient fields.

\subsection{Gradients in $B_z$}

The actively-shielded coil, discussed extensively in Sec.\ \ref{sec:Coil}, produces the highly homogeneous $B_z$ throughout the precession volume. Normally, the halves of both the inner and outer coils in this system, to either side of the xy plane, are connected in series so that the same current flows through both halves. We have relays that enable us to reverse the current direction on one half of both coils, whereupon the main coils starts producing a gradient $\dB{z}{z}$. Given Gauss' Law, of course, this means that we are also producing gradients $\dB{x}{x}$ and $\dB{y}{y}$ such that 
\begin{align}
\dB{z}{z}  = - \dB{x}{x}- \dB{y}{y}. 
\label{dBzdz}
\end{align}
The geometry of the actively-shielded coil is roughly translational-symmetric along $x$ coordinate for the central region of the coil, thus, we expect the $\dB{z}{z}$ coil to primarily produce $\dB{y}{y}$, while little $\dB{x}{x}$ would be generated alongside. Explicit calculations give
\begin{align}
\dB{x}{x} &= - 0.06 \, \dB{z}{z}\\
\dB{y}{y} &= - 0.94 \, \dB{z}{z},\\
\end{align}

they are confirmed by the measurements as shown in Table \ref{tab:gradient_components}, and they satisfy Eq.\ \ref{dBzdz}.  
Within the precession volume, the calculated gradient varies from the nominal value at the center of this volume by about 6.9 \% at maximum.  The good agreement between the calculated and measured gradients is shown in Fig.\ \ref{fig:Gradients}a.

\section{Q State Magnetic Dipole Moment}
\label{sec:Q state dipole moment} 

The magnetic dipole moment of the Q state has previously been measured from the Zeeman splitting under the application of a large magnetic field \cite{Wu2020}. To obtain a more precise measurement, we measure the Q state magnetic dipole moment using a spin precession scheme in the ACME III beamline. The molecules are prepared in an incoherent mixture of all Q state m-sublevels as described in \ref{sec:comag}. No transverse preparation laser is applied. Instead, an 1196 nm y-polarized laser resonant with the Q$\rightarrow$C transition is sent from the downstream window at the end of the beamline. The laser is pulsed on for 3 ms while the molecules travel through the interaction region and prepares them to an initial state $(\ket{Q, M=+1}+\ket{Q, M=-1}) / \sqrt{2}$. The end of the laser pulse is chosen to coincide with the first molecules arriving at the readout location. The phase is computed from the asymmetry as described for the co-magnetometry measurement as a function of time after the preparation pulse is shut off. The phase measurement is taken once with an applied $B_z=+|B|$ and once with an applied $B_z=-|B|$ for $B\in \{50, 62.5, 75, 87.5, 100\}\  \mu$G. The differential phase evolution per unit time between the reversed B field states is given by

\begin{equation}
\frac{d\phi}{dt}(B_z=+|B|)-\frac{d\phi}{dt}(B_z=-|B|)=\frac{2g_Q\mu_BR(J, M, \Omega)}{\hbar}
\end{equation}

where $\mu_B$ is the Bohr magneton and $R(J, M, \Omega)$ is a rotational constant. We solve this equation for $g_Q$ and compute the value at each field magnitude to confirm consistency. 

\begin{figure}
    \centering
     \includegraphics[width=\columnwidth]{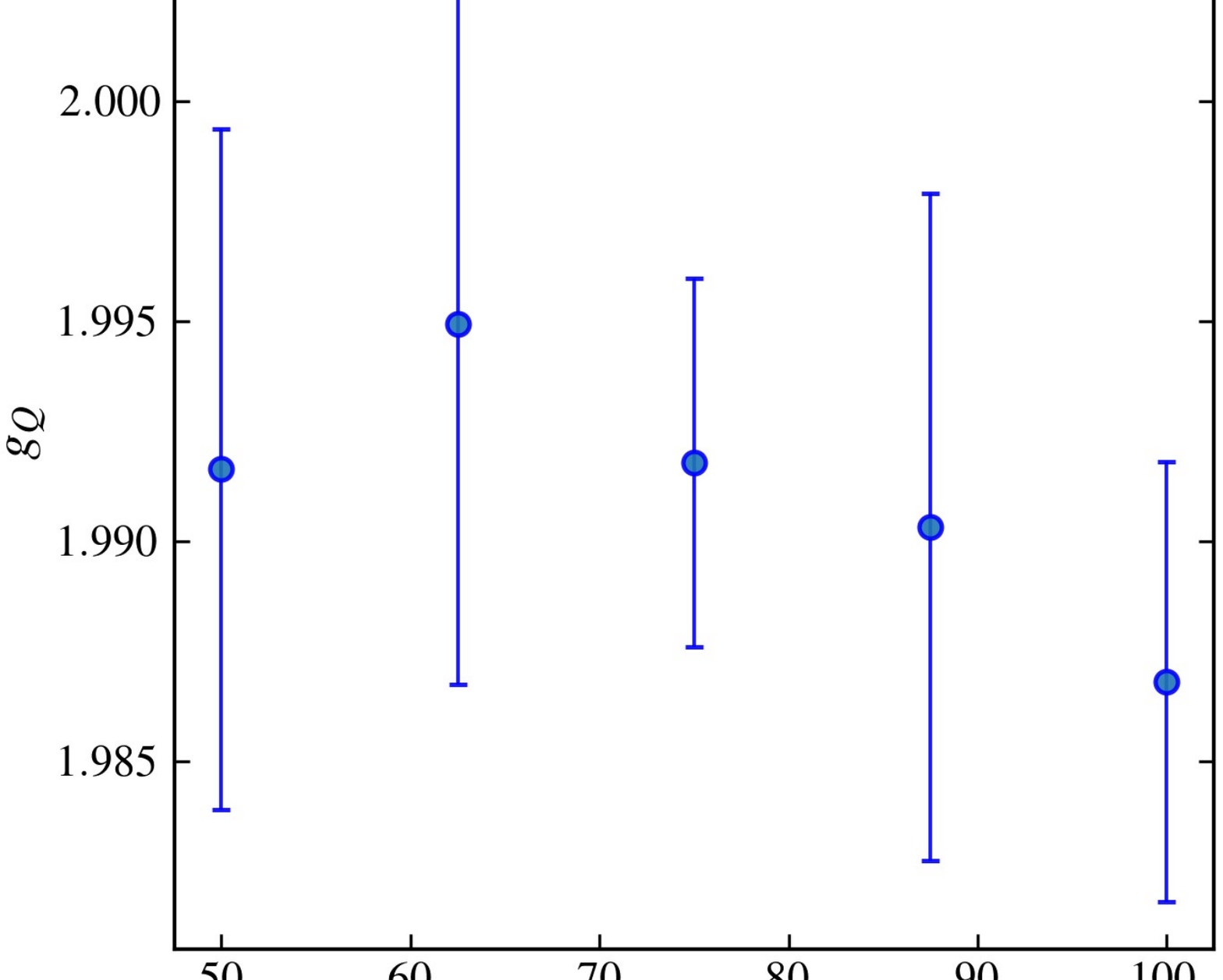}
    \caption{Consistency check of the measured g-factor at several different magnitudes of reversing applied magnetic fields $B_z$. The plotted error bar includes only the statistical uncertainty.}
    \label{fig:Q g factor}
\end{figure}

The uncertainty is limited by three factors. First, our absolute calibration of the applied field is limited to 0.5\% from the specified accuracy of the flux-gate magnetometers. Additionally, we have 0.18\% variation of the applied magnetic field along the spin precession region. Finally, the measurement can only be performed in one $N=M\Omega E$ state, so the $N$-correlated g-factor cannot be measured. To leading order, this contribution to the g factor may be written:
\begin{equation}
g_Q^N=\frac{d_QEg}{B_Q}\alpha(J, \Omega)
\end{equation}
where $d_Q$ is the electric dipole moment of the ThO Q state, $E$ is the applied electric field, $B_Q$ is the rotational constant of the Q state, and $\alpha(J, \Omega)$ is a rotational factor. We estimate the size of this effect to be 0.05\% and include this as an additional systematic error. Adding the uncertainties in quadrature, we find $g_Q=1.991\pm0.012$ which is in good agreement with, but more precise than, the previous measurement.

\section{H State Magnetic Dipole Moment}
\label{sec:H state dipole moment} 

In a precession time $\tau$, the ThO $H$ state (with an electric dipole moment $d$ and a magnetic dipole moment $\mu$) evolves by a measurable phase $\phi$ given by   
\begin{equation}
\hbar \phi =- (d E + \mu B)\tau.
\end{equation}
The precession time $\tau$ must be measured to make it possible to determine $d$ from measured changes in the precession phase as $E$ is changed.  Changes in the precession phase $\phi$ as the magnetic field is changed can be used to determine $\tau$ once the magnetic moment $\mu = g \mu_B$ is measured precisely enough. 
To measure g, the change in precession phase $\phi$ is measured as prepared H-states travel a distance $L=50\pm 0.5$ cm,  within a precisely calibrated magnetic field produced as described in this report. The precession time is then $\tau=L/v$, where the measured velocity described below is typically about 180 m/s.

\begin{figure}
    \centering
     \includegraphics[width=\columnwidth]{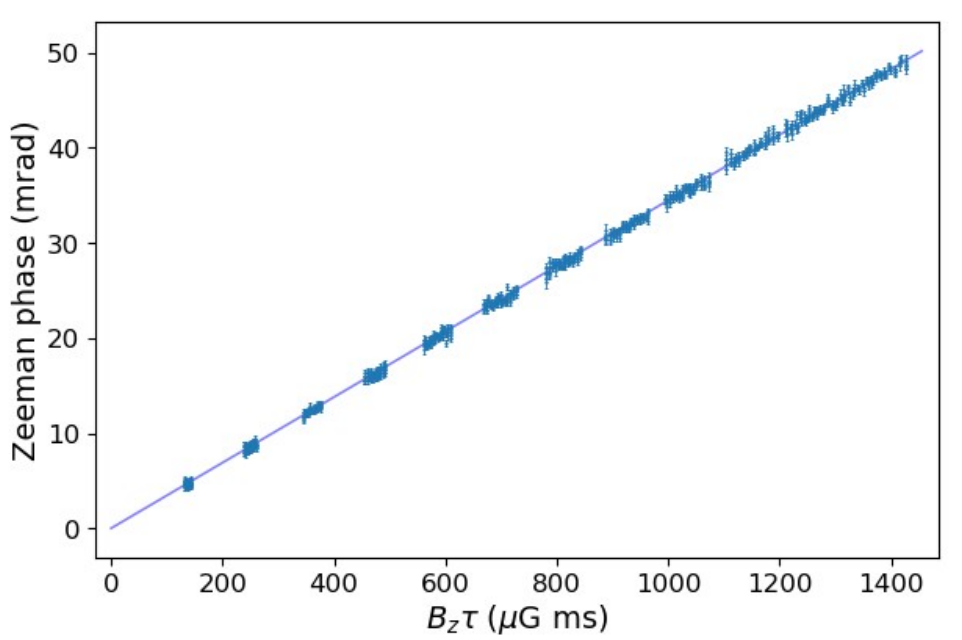}
    \caption{Fit of the Zeeman precession phase as a function of the product $B_z \tau$}
    \label{fig:H g factor}
\end{figure}

After rotational cooling and electrostatic lensing, molecules are optically pumped back to $\ket{X, J=0}$. In the interaction region, an electric field of 80 V/cm and reversible magnetic fields of various magnitudes are applied. The molecules are then transferred via STIRAP through the intermediate C state to $(\ket{H, M=+1, N}+\ket{H, M=-1, N})/\sqrt{2}$. The molecules are illuminated by a laser with two frequency components, one resonant with the $\ket{H, |M|=1,  N}\rightarrow\ket{I, M=0, P}$ transitions for excited state parity $P=\pm1$. These frequency components are separated by 91 MHz and produced by two separate acousto-optic modulators (AOM) coupled to the same polarization maintaining fiber. For the same laser polarization, changing the excited state parity rotates the bright and dark states of the laser-molecule interaction by 90 degrees. Thus, when both parity drives are on, all molecules are in a bright state of one the frequencies and are optically pumped out of H. To allow a brief pulse of molecules to pass through the state preparation laser, the P=+1 frequency is shut off for $T=$ 200 $\mu$s. The timing is controlled by a delay generator that also produces the 50 Hz trigger for data taking. When only one parity drive is on, each molecule is prepared into the dark state of the laser.

We model the molecule population prepared as a square pulse in time with a Gaussian distribution of velocity. In this case the population observed at readout has a  time dependence that is determined by the integral:

\begin{equation}
\begin{split}
S(t) =\frac{1}{2T\sqrt{2\pi\sigma^2_v}}\int_0^\infty\Theta(t-\frac{L}{v}) \\ \times \Theta(T-t+\frac{L}{v})e^{-\frac{(v-v_0)^2}{2\sigma_v^2}}dv
\label{eq:signal}
\end{split}
\end{equation}
where $\Theta$ is the Heaviside step function. Due to the velocity dispersion of the molecular beam, the precession time varies by $\sim$ 5 percent over the detected pulse. We fit the detected fluorescence according to \eqref{eq:signal} to extract the average velocity $v_0$ and standard deviation of velocity $\sigma_v$. The precession time throughout the detected beam can then be calculated using 

\begin{equation}
\begin{split}
\tau(t) =\frac{1}{2T\sqrt{2\pi\sigma^2_v}}\int_0^\infty\frac{L}{v}\Theta(t-\frac{L}{v}) \\ \times \Theta(T-t+\frac{L}{v})e^{-\frac{(v-v_0)^2}{2\sigma_v^2}}dv
\end{split}
\end{equation}

The molecule precession phase is determined with the polarization switching scheme outlined in \ref{sec:comag} under a reversing, applied $B_z$ field at 12 different magnitudes ranging from 50 $\mu$G to 500 $\mu$G. The measurement is performed in both N states and the precession phase is averaged over the two. The phase that reverses with the applied magnetic field is fit to a line with respect to $B_z \tau$, shown in \ref{fig:H g factor}, and the g-factor can be computed from the slope S:

\begin{equation}
g_H=-\frac{S}{\mu_B R(J, M, \Omega)}
\end{equation}
where $R(J, M, \Omega)=1/2$ is a rotational factor. As in the Q state measurement, we include a 0.5\% systematic error from the specified absolute accuracy of the flux-gate magnetometers and a 0.18\% systematic error due to the variation of the applied magnetic field along the spin precession region. Additionally, as a conservative estimate of the maximum error in the precession time, we include a systematic uncertainty from the standard deviation of fit precession times, 2.4\%. Combined in quadrature with the statistical uncertainty, we find a measured g-factor $g_H=-0.0078(2)$. This is in tension with the previous most precise measurement which found $g_H=-0.0088(1)$ \cite{previousgH}. The cause of the discrepancy is not known. However, we find several independent observations support this updated measurement. First, the precession time as measured by the Zeeman phase in the H and Q state agree within uncertainty, provided the new value is used. Second, a measurement of the electric field was done using microwave spectroscopy between the H, J=1 and H, J=2 manifolds. The $\sim$ 40 GHz microwaves are sent into the interaction region from the downstream window, similar to the state preparation laser in \ref{sec:Q state dipole moment}. In addition to the intended transition, we find that reflections of the microwaves from a molecular beam collimator upstream also cause rotational transitions, separated by 45 to 50 kHz. This matches the Doppler shift expected for a molecular beam traveling at 180 m/s, consistent with the precession time measured with the newest measurement of $g_H$.

\bibliography{shield_paper_refs}

@article{AcmeEdm2014,
  title = {Order of Magnitude Smaller Limit on the Electric Dipole Moment of the Electron.},
  author = {Baron, J. and Campbell, W. C. and DeMille, D. and Doyle, J. M. and
Gabrielse, G. and Gurevich, Y. V. and Hess, P. W. and Hutzler, N. R. and
Kirilov, E. and Kozyryev, I. and O'Leary, B. R. and Panda, C. D. and
Parsons, M. F. and Petrik, E. S. and Spaun, B. and Vutha, A. C. and West, A. D.},
  year = {2014},
  journal = {Science (New York, N.Y.)},
  volume = {343},
  number = {6168},
  pages = {269--72},
  issn = {1095-9203},
  doi = {10.1126/science.1248213},
  pmid = {24356114}
}

@article{AcmeEdm2018,
  title = {Improved Limit on the Electric Dipole Moment of the Electron},
  author = {Andreev, V. and Ang, D. G. and DeMille, D. and Doyle, J. M. and Gabrielse, G. and Haefner, J. and Hutzler, N. R. and Lasner, Z. and Meisenhelder, C. and O'Leary, B. R. and Panda, C. D. and West, A. D. and West, E. P. and Wu, X.},
  year = {2018},
  journal = {Nature},
  volume = {562},
  number = {7727},
  pages = {355--360},
  issn = {0028-0836},
  doi = {10.1038/s41586-018-0599-8}
}

@article{AcmeEdmDetails2016,
  title = {Methods, Analysis, and the Treatment of Systematic Errors for the Electron Electric Dipole Moment Search in Thorium Monoxide},
  author = {Baron, J.  and Campbell, W. C. and DeMille, D. and Doyle, J. M. and Gabrielse, G. and Gurevich, Y. V. and Hess, P. W. and Hutzler, N. R. and Kirilov, E. and Kozyryev, I. and O'Leary, B. R. and Panda, C. D. and Parsons, M. F. and Spaun, B. and Vutha, A. C. and West, A. D. and West, E. P.},
  year = {2016},
  journal = {New Journal of Physics},
  volume = {19},
  pages = {073029},
  publisher = {{IOP Publishing}},
  doi = {10.1088/1367-2630/aa708e}
}

@article{Reece2019,
  title = {Interpreting the Electron {{EDM}} Constraint},
  author = {Cesarotti,  C. and Lu,  Q. and Nakai,  Y. and Parikh,  A. and Reece,  M.},
  year = {2019},
  month = may,
  journal = {Journal of High Energy Physics},
  volume = {2019},
  number = {5},
  pages = {59},
  issn = {1029-8479},
  doi = {10.1007/JHEP05(2019)059}
}

@phdthesis{PandaThesis2019,
  title = {Order of Magnitude Improved Limit on the Electric Dipole Moment of the Electron},
  author = {Panda, C. D.},
  year = {2019},
  school = {Harvard University}
}

@article{Arpaia2021,
  doi = {10.1016/j.nima.2020.164904},
  url = {https://doi.org/10.1016/j.nima.2020.164904},
  year = {2021},
  month = feb,
  publisher = {Elsevier {BV}},
  volume = {988},
  pages = {164904},
  author = {Arpaia, P.  and Burrows, P. N. and Buzio,  M. and Gohil, C. and Pentella, M. and Schulte D.},
  title = {Magnetic characterization of Mumetal{\textregistered} for passive shielding of stray fields down to the nano-Tesla level},
  journal = {Nuclear Instruments and Methods in Physics Research Section A: Accelerators,  Spectrometers,  Detectors and Associated Equipment}
}

@techreport{crawfordmartinbfield,
  title = {A Double Cosine Theta Coil Prototype},
  author = {Martin, E. and Crawford, C.},
  year = {2010},
  institution = {{University of Kentucky}},
  url = {https://web.archive.org/web/20230209004845/https://www.pa.uky.edu/~crawford/pub/coscoil4.pdf}
}

@techreport{crawfordshinbfield,
  title = {A method for designing coils
with arbitrary fields},
  author = {Crawford, C. and Shin, Y.},
  year = {2009},
  institution = {{University of Kentucky}},
  url = {http://web.archive.org/web/20230209004834/https://www.pa.uky.edu/~crawford/pub/dsctc.pdf}
}

@book{jackson1998classical,
  title = {Classical Electrodynamics},
  author = {Jackson, J.D.},
  year = {1998},
  publisher = {{Wiley}},
  isbn = {978-0-471-30932-1},
  lccn = {97046873}
}

@article{Preece1971,
  doi = {10.1109/tmag.1971.1067192},
  url = {https://doi.org/10.1109/tmag.1971.1067192},
  year = {1971},
  month = sep,
  publisher = {Institute of Electrical and Electronics Engineers ({IEEE})},
  volume = {7},
  number = {3},
  pages = {554--557},
  author = {I. Preece and R. Thomas},
  title = {The effects of various stresses on mumetal: 77 percent Ni-14 percent Fe-5 percent Cu-4 percent Mo},
  journal = {{IEEE} Transactions on Magnetics}
}

@inbook{MagneticShieldTextbook_2013, place={Cambridge}, title={Magnetic shielding}, DOI={10.1017/CBO9780511846380.013}, booktitle={Optical Magnetometry}, publisher={Cambridge University Press}, author={Yashchuk, V. V. and Lee, S.-K. and Paperno, E.}, editor={Budker, Dmitry and Jackson Kimball, Derek F.Editors}, year={2013}, pages={225–248}}

@article{Chubar1998,
  title = {A three-dimensional magnetostatics computer code for insertion devices},
  volume = {5},
  ISSN = {0909-0495},
  url = {http://dx.doi.org/10.1107/S0909049597013502},
  DOI = {10.1107/s0909049597013502},
  number = {3},
  journal = {Journal of Synchrotron Radiation},
  publisher = {International Union of Crystallography (IUCr)},
  author = {Chubar,  O. and Elleaume,  P. and Chavanne,  J.},
  year = {1998},
  month = may,
  pages = {481–484}
}

@inproceedings{inproceedings_BMSR,
author = {Bork, J. and Hahlbohm, H. D and Klein, R. and Schnabel, A.},
year = {2001},
month = {01},
pages = {},
title = {The 8-layered magnetically shielded room of the PTB: Design and Construction},
journal = {Proc. Biomag}
}

@article{AcmeLifetime2022,
  title = {Measurement of the {$H^{3}$}{{$\Delta$}}{$_1$} Radiative Lifetime in {{ThO}}},
  author = {Ang, D. G. and Meisenhelder, C. and Panda, C. D. and Wu, X. and DeMille, D. and Doyle, J. M. and Gabrielse, G.},
  year = {2022},
  month = aug,
  journal = {Physical Review A: Atomic, Molecular, and Optical Physics},
  volume = {106},
  number = {2},
  pages = {022808},
  publisher = {{American Physical Society}},
  doi = {10.1103/PhysRevA.106.022808}
}

@article{Denis2016eeff,
author = {Denis, M. and Fleig, T. },
title = {In search of discrete symmetry violations beyond the standard model: Thorium monoxide reloaded},
journal = {The Journal of Chemical Physics},
volume = {145},
number = {21},
pages = {214307},
year = {2016},
doi = {10.1063/1.4968597},
URL = {https://doi.org/10.1063/1.4968597}
}

@phdthesis{HutzlerThesis2014,
  title = {A New Limit on the Electron Electric Dipole Moment: {{Beam}} Production, Data Interpretation, and Systematics},
  author = {Hutzler, N. R.},
  year = {2014},
  number = {February},
  pages = {1--322},
  school = {Harvard University}
}

@phdthesis{LasnerThesis2019,
  title = {Order-of-Magnitude-Tighter Bound on the Electron Electric Dipole Moment},
  author = {Lasner, Z.},
  year = {2019},
  school = {Yale University}
}

@article{masudasipm,
  title = {Suppression of the Optical Crosstalk in a Multi-Channel Silicon Photomultiplier Array},
  author = {Masuda, T. and Ang, D. G. and Hutzler, N. R. and Meisenhelder, C. and Sasao, N. and Uetake, S. and Wu, X. and DeMille, D. and Gabrielse, G. and Doyle, J. M. and Yoshimura, K.},
  year = {2021},
  month = may,
  journal = {Optics Express},
  volume = {29},
  number = {11},
  pages = {16914--16926},
  publisher = {{OSA}},
  doi = {10.1364/OE.424460}
}

@article{AcmeSipm2023,
author = { Masuda, T. and Hiramoto, A and Ang, D. G.  and  Meisenhelder, C. and Panda, C. D. and Sasao, N. and Uetake, S. and Wu,  X. and DeMille, D. and Doyle, J. M. and Gabrielse, G. and Yoshimura, K.},
journal = {Opt. Express},
number = {2},
pages = {1943--1957},
publisher = {Optica Publishing Group},
title = {High-sensitivity low-noise photodetector using a large-area silicon photomultiplier},
volume = {31},
month = {Jan},
year = {2023},
url = {https://opg.optica.org/oe/abstract.cfm?URI=oe-31-2-1943},
doi = {10.1364/OE.475109},
}

@article{Reece2017,
  title = {Electric Dipole Moments in Natural Supersymmetry},
  author = {Nakai, Y. and Reece, M.},
  year = {2017},
  month = aug,
  journal = {Journal of High Energy Physics},
  volume = {2017},
  number = {8},
  pages = {31},
  issn = {1029-8479},
  doi = {10.1007/JHEP08(2017)031},
  langid = {english}
}

@article{AcmeShotNoise2019,
  title = {Attaining the Shot-Noise-Limit in the {{ACME}} Measurement of the Electron Electric Dipole Moment},
  author = {Panda, C. D. and Meisenhelder, C. and Verma, M. and Ang, D. G. and Chow, J. and Lasner, Z. and Wu, X. and DeMille, D. and Doyle, J. M. and Gabrielse, G.},
  year = {2019},
  month = nov,
  journal = {Journal of Physics B: Atomic, Molecular and Optical Physics},
  volume = {52},
  number = {23},
  pages = {235003},
  publisher = {{IOP Publishing}},
  doi = {10.1088/1361-6455/ab4a61}
}

@phdthesis{AngThesis2023,
  title = {Progress on an Improved Measurement of the Electron Electric Dipole Moment},
  author = {Ang, D. G.},
  year = {2023},
  school = {Harvard University}
}

@phdthesis{MeisenhelderThesis2023,
  title = {Advances in the Measurement of the Electron Electric Dipole Moment},
  author = {Meisenhelder, C.},
  year = {2023},
  school = {Harvard University}
}

@article{Petrov2014,
  title = {Zeeman Interaction in {{ThO H}}{$^{3}$}{{$\Delta$}}{$_1$} for the Electron Electric-Dipole-Moment Search},
  author = {Petrov, A. N. and Skripnikov, L. V. and Titov, A. V. and Hutzler, N. R. and Hess, P. W. and O'Leary, B. R. and Spaun, B. and DeMille, D. and Gabrielse, G. and Doyle, J. M.},
  year = {2014},
  journal = {Physical Review A: Atomic, Molecular, and Optical Physics},
  volume = {89},
  number = {6},
  pages = {062505},
  issn = {1050-2947},
  doi = {10.1103/PhysRevA.89.062505}
}

@article{Skripnikov2016,
author = {Skripnikov,L. V. },
title = {Combined 4-component and relativistic pseudopotential study of {ThO} for the electron electric dipole moment search},
journal = {The Journal of Chemical Physics},
volume = {145},
number = {21},
pages = {214301},
year = {2016},
doi = {10.1063/1.4968229},
URL = {https://doi.org/10.1063/1.4968229}
}

@phdthesis{SpaunThesis2014,
  title = {A Ten-Fold Improvement to the Limit of the Electron Electric Dipole Moment},
  author = {Spaun, Benjamin Norman},
  year = {2014},
  number = {May},
  school = {Harvard University}
}

@article{Hiramoto2022sipm,
  title = {{{SiPM}} Module for the {{ACME III}} Electron {{EDM}} Search},
  author = {Hiramoto, A. and Masuda, T. and Ang, D.G. and Meisenhelder, C. and Panda, C. and Sasao, N. and Uetake, S. and Wu, X. and Demille, D. and Doyle, J.M. and Gabrielse, G. and Yoshimura, K.},
  year = {2023},
  journal = {Nuclear Instruments and Methods in Physics Research Section A: Accelerators, Spectrometers, Detectors and Associated Equipment},
  volume = {1045},
  pages = {167513},
  issn = {0168-9002},
  doi = {10.1016/j.nima.2022.167513}
}

@article{hanCryogenicBufferGas2026,
	title = {Cryogenic buffer gas beam source with in situ ablation target replacement},
	volume = {1},
	url = {https://link.aps.org/doi/10.1103/3qhm-yr7d},
	doi = {10.1103/3qhm-yr7d},
	urldate = {2026-07-24},
	journal = {APS Open Science},
	author = {Han, Z. and Lasner, Z. and Diver, C. and Hu, P. and Masuda, T. and Wu, X. and Hiramoto, A. and Watts, M. and Uetake, S. and Yoshimura, K. and Fan, X. and Gabrielse, G. and Doyle, J. M. and DeMille, D.},
	month = may,
	year = {2026},
	note = {Publisher: American Physical Society},
	pages = {000016},
}

@article{Wu2020,
  title = {The Metastable {{Q}} {$^{3}$}{{$\Delta$}}{$_2$} State of {{ThO}}: A New Resource for the {{ACME}} Electron {{EDM}} Search},
  author = {Wu, X. and Han, Z. and Chow, J. and Ang, D. G. and Meisenhelder, C. and Panda, C. D. and West, E. P. and Gabrielse, G. and Doyle, J. M. and DeMille, D.},
  year = {2020},
  month = feb,
  journal = {New Journal of Physics},
  volume = {22},
  number = {2},
  pages = {023013},
  publisher = {{IOP Publishing}},
  doi = {10.1088/1367-2630/ab6a3a}
}

@article{AcmeLens2022,
  title = {Electrostatic Focusing of Cold and Heavy Molecules for the {{ACME}} Electron {{EDM}} Search},
  author = {Wu, X. and Hu, P. and Han, Z. and Ang, D. G. and Meisenhelder, C. and Gabrielse, G. and Doyle, J. M. and DeMille, D.},
  year = {2022},
  month = jul,
  journal = {New Journal of Physics},
  volume = {24},
  number = {7},
  pages = {073043},
  publisher = {{IOP Publishing}},
  doi = {10.1088/1367-2630/ac8014}
}

@article{jilaedm2023,
  author  = {Roussy, T. S. and Caldwell, L. and Wright, T. and Cairncross, W. B. and Shagam, Y. and Ng, K. B. and Schlossberger, N. and Park, S. Y. and Wang, A. and Ye, J. and Cornell, E. A.},
  title   = {An improved bound on the electron's electric dipole moment},
  journal = {Science},
  volume  = {381},
  number  = {6653},
  pages   = {46--50},
  year    = {2023},
  doi     = {10.1126/science.adg4084},
  url     = {https://www.science.org/doi/abs/10.1126/science.adg4084}
}

@article{neutronEDMShield2014PSI,
    author = {Afach, S. and Bison, G. and Bodek, K. and Burri, F. and Chowdhuri, Z. and Daum, M. and Fertl, M. and Franke, B. and Grujic, Z. and Hélaine, V. and Henneck, R. and Kasprzak, M. and Kirch, K. and Koch, H.-C. and Kozela, A. and Krempel, J. and Lauss, B. and Lefort, T. and Lemière, Y. and Meier, M. and Naviliat-Cuncic, O. and Piegsa, F. M. and Pignol, G. and Plonka-Spehr, C. and Prashanth, P. N. and Quéméner, G. and Rebreyend, D. and Roccia, S. and Schmidt-Wellenburg, P. and Schnabel, A. and Severijns, N. and Voigt, J. and Weis, A. and Wyszynski, G. and Zejma, J. and Zenner, J. and Zsigmond, G.},
    title = "{Dynamic stabilization of the magnetic field surrounding the neutron electric dipole moment spectrometer at the Paul Scherrer Institute}",
    journal = {Journal of Applied Physics},
    volume = {116},
    number = {8},
    pages = {084510},
    year = {2014},
    month = {08},
    issn = {0021-8979},
    doi = {10.1063/1.4894158},
    url = {https://doi.org/10.1063/1.4894158},
}

@article{neutronEDMMagneticShield2017ORNL,
title = {Cryogenic magnetic coil and superconducting magnetic shield for neutron electric dipole moment searches},
journal = {Nuclear Instruments and Methods in Physics Research Section A: Accelerators, Spectrometers, Detectors and Associated Equipment},
volume = {862},
pages = {36-48},
year = {2017},
issn = {0168-9002},
doi = {https://doi.org/10.1016/j.nima.2017.05.005},
url = {https://www.sciencedirect.com/science/article/pii/S016890021730534X},
author = {S. Slutsky and C.M. Swank and A. Biswas and R. Carr and J. Escribano and B.W. Filippone and W.C. Griffith and M. Mendenhall and N. Nouri and C. Osthelder and A. {Pérez Galván} and R. Picker and B. Plaster}
}

@article{PbO3Delta1Proposal2001,
    author = {DeMille, D. and Bay, F. and Bickman, S. and Kawall, D. and Hunter, L. and Krause, D., Jr. and Maxwell, S. and Ulmer, K.},
    title = "{Search for the electric dipole moment of the electron using metastable PbO}",
    journal = {AIP Conference Proceedings},
    volume = {596},
    number = {1},
    pages = {72-83},
    year = {2001},
    month = {11},
    issn = {0094-243X},
    doi = {10.1063/1.1426795},
    url = {https://doi.org/10.1063/1.1426795},
}

@article{thiel_demagnetization_2007,
	title = {Demagnetization of magnetically shielded rooms},
	volume = {78},
	issn = {0034-6748, 1089-7623},
	url = {http://aip.scitation.org/doi/10.1063/1.2713433},
	doi = {10.1063/1.2713433},
	number = {3},
	urldate = {2023-03-14},
	journal = {Review of Scientific Instruments},
	author = {Thiel, F. and Schnabel, A. and Knappe-Grüneberg, S. and Stollfuß, D. and Burghoff, M.},
	month = mar,
	year = {2007},
	pages = {035106},
}

@article{Holmes2022,
  title = {A lightweight magnetically shielded room with active shielding},
  volume = {12},
  ISSN = {2045-2322},
  url = {http://dx.doi.org/10.1038/s41598-022-17346-1},
  DOI = {10.1038/s41598-022-17346-1},
  number = {1},
  journal = {Scientific Reports},
  publisher = {Springer Science and Business Media LLC},
  author = {Holmes,  N. and Rea,  M. and Chalmers,  J. and Leggett,  J. and Edwards,  L. J. and Nell,  P. and Pink,  S. and Patel,  P. and Wood,  J. and Murby,  N. and Woolger,  D. and Dawson,  E. and Mariani,  C. and Tierney,  T. M. and Mellor,  S. and O’Neill,  G. C. and Boto,  E. and Hill,  R. M. and Shah,  V. and Osborne,  J. and Pardington,  R. and Fierlinger,  P. and Barnes,  G. R. and Glover,  P. and Brookes,  M. J. and Bowtell,  R.},
  year = {2022},
  month = aug 
}

@ARTICLE{ActiveShieldingMRI2,
  author={Ishiyama, A. and Hondoh, M. and Ishida, N. and Onuki, T.},
  journal={IEEE Transactions on Magnetics}, 
  title={Optimal design of MRI magnets with magnetic shielding}, 
  year={1989},
  volume={25},
  number={2},
  pages={1885-1888},
  doi={10.1109/20.92673}}

@ARTICLE{ActiveShieldingMRI1,
  author={Hawksworth, D. and McDougall, I. and Bird, J. and Black, D.},
  journal={IEEE Transactions on Magnetics}, 
  title={Considerations in the design of MRI magnets with reduced stray fields}, 
  year={1987},
  volume={23},
  number={2},
  pages={1309-1314},
  doi={10.1109/TMAG.1987.1065051}}

@article{CrawfordRSI2021,
    author = {Crawford, C. B.},
    title = "{The physical meaning of the magnetic scalar potential and its use in the design of hermetic electromagnetic coils}",
    journal = {Review of Scientific Instruments},
    volume = {92},
    number = {12},
    pages = {124703},
    year = {2021},
    month = {12},
    issn = {0034-6748},
    doi = {10.1063/5.0063054},
    url = {https://doi.org/10.1063/5.0063054},
}

@article{FlavorProbes2022,
    author = {Bauer, M.  and Neubert, M.  and Renner, S. and Schnubel, M.  and Thamm, A. },
    title = "{Flavor probes of axion-like particles}",
    journal = {Journal of High Energy Physics},
   pages = {56},
    year = {2022},
    url = {https://link.springer.com/article/10.1007/JHEP09(2022)056}, 
}

@article{Baryogenesis2021,
  title = {Baryogenesis from the weak scale to the grand unification scale},
  author = {B\"odeker, D. and Buchm\"uller, W.},
  journal = {Rev. Mod. Phys.},
  volume = {93},
  issue = {3},
  pages = {035004},
  numpages = {45},
  year = {2021},
  month = {Aug},
  publisher = {American Physical Society},
  doi = {10.1103/RevModPhys.93.035004},
  url = {https://link.aps.org/doi/10.1103/RevModPhys.93.035004}
}

@article{COMSOL,
  title={Introduction to COMSOL multiphysics{\textregistered}},
  author={Multiphysics, COMSOL},
  journal={COMSOL Multiphysics® v. 6.2. www.comsol.com. COMSOL AB, Stockholm, Sweden.},
}

@article{neutronEDMShield2015TUM,
    author = {Altarev, I. and Bales, M. and Beck, D. H. and Chupp, T. and Fierlinger, K. and Fierlinger, P. and Kuchler, F. and Lins, T. and Marino, M. G. and Niessen, B. and Petzoldt, G. and Schläpfer, U. and Schnabel, A. and Singh, J. T. and Stoepler, R. and Stuiber, S. and Sturm, M. and Taubenheim, B. and Voigt, J.},
    title = "{A large-scale magnetic shield with 106 damping at millihertz frequencies}",
    journal = {Journal of Applied Physics},
    volume = {117},
    number = {18},
    pages = {183903},
    year = {2015},
    month = {05},
    issn = {0021-8979},
    doi = {10.1063/1.4919366},
    url = {https://doi.org/10.1063/1.4919366},
}

@article{neutronEDMTUMMagneticShield2014,
    author = {Altarev, I. and Babcock, E. and Beck, D. and Burghoff, M. and Chesnevskaya, S. and Chupp, T. and Degenkolb, S. and Fan, I. and Fierlinger, P. and Frei, A. and Gutsmiedl, E. and Knappe-Grüneberg, S. and Kuchler, F. and Lauer, T. and Link, P. and Lins, T. and Marino, M. and McAndrew, J. and Niessen, B. and Paul, S. and Petzoldt, G. and Schläpfer, U. and Schnabel, A. and Sharma, S. and Singh, J. and Stoepler, R. and Stuiber, S. and Sturm, M. and Taubenheim, B. and Trahms, L. and Voigt, J. and Zechlau, T.},
    title = "{A magnetically shielded room with ultra low residual field and gradient}",
    journal = {Review of Scientific Instruments},
    volume = {85},
    number = {7},
    pages = {075106},
    year = {2014},
    month = {07},
    issn = {0034-6748},
    doi = {10.1063/1.4886146},
    url = {https://doi.org/10.1063/1.4886146},
}

@misc{CoNeticAA,
  author = {{Magnetic Shield Corp.}},
  title = {{Co-NETIC®  AA Perfection Sheet}},
  howpublished = {\url{https://www.magnetic-shield.com/co-netic-aa-perfection-sheet/}},
  note         = {Accessed August 4, 2026}
}

@article{Gabrielse1988,
  title = {Self-shielding superconducting solenoid systems},
  volume = {63},
  ISSN = {1089-7550},
  url = {http://dx.doi.org/10.1063/1.340416},
  DOI = {10.1063/1.340416},
  number = {10},
  journal = {Journal of Applied Physics},
  publisher = {AIP Publishing},
  author = {Gabrielse,  G. and Tan,  J.},
  year = {1988},
  month = may,
  pages = {5143–5148}
}

@article{Gabrielse1991,
  title = {A superconducting solenoid system which cancels fluctuations in the ambient magnetic field},
  volume = {91},
  ISSN = {0022-2364},
  url = {http://dx.doi.org/10.1016/0022-2364(91)90382-4},
  DOI = {10.1016/0022-2364(91)90382-4},
  number = {3},
  journal = {Journal of Magnetic Resonance (1969)},
  publisher = {Elsevier BV},
  author = {Gabrielse,  G. and Tan,  J. and Clateman,  P.  and Orozco,  L.A. and Rolston,  S.L. and Tseng,  C.H. and Tjoelker,  R.L.},
  year = {1991},
  month = feb,
  pages = {564–572}
}

@article{previousgH,
  title = {Shot-noise-limited spin measurements in a pulsed molecular beam},
  author = {Kirilov, E. and Campbell, W. C. and Doyle, J. M. and Gabrielse, G. and Gurevich, Y. V. and Hess, P. W. and Hutzler, N. R. and O'Leary, B. R. and Petrik, E. and Spaun, B. and Vutha, A. C. and DeMille, D.},
  journal = {Phys. Rev. A},
  volume = {88},
  issue = {1},
  pages = {013844},
  numpages = {9},
  year = {2013},
  month = {Jul},
  publisher = {American Physical Society},
  doi = {10.1103/PhysRevA.88.013844},
  url = {https://link.aps.org/doi/10.1103/PhysRevA.88.013844}
}

@article{StaticVQuasi1,
  title={Limits of low magnetic field environments in magnetic shields},
  author={Sun, Z. and Fierlinger, P. and Han, J. and Li, Liyi and Liu, T. and Schnabel, A. and Stuiber, S. and Voigt, J.},
  journal={IEEE Transactions on Industrial Electronics},
  volume={68},
  number={6},
  pages={5385--5395},
  year={2020},
  publisher={IEEE},
  url = {https://ieeexplore.ieee.org/document/9075432}
}

@article{StaticVQuasi2,
  title = {Experimental studies on the performance of magnetic shields under different magnetization conditions},
  author = {Yang, J. and Zhang, X. and Shi, M. and Yuan, S. and Zhang, L. and Wang, L. and Han, B.},
  journal = {Journal of Physics D: Applied Physics},
  volume = {56},
  number = {21},
  pages = {215001},
  year = {2023},
  publisher = {IOP Publishing},
  url = {https://iopscience.iop.org/article/10.1088/1361-6463/acc412}
}

\end{document}